\documentclass{article}

\usepackage{arxiv}

\usepackage[T1]{fontenc}
\usepackage[utf8]{inputenc}

\usepackage{amsmath}
\usepackage{amssymb}
\usepackage{graphicx}
\usepackage{booktabs}
\usepackage{microtype}

\usepackage[round]{natbib}
\usepackage[hyphens]{url}
\usepackage[hidelinks]{hyperref}

\graphicspath{{Figures/}}

\title{When Does Latent Communication Pay? A Causal Audit of Relayed KV
Caches in Multi-Agent LLMs}
\author{
  Jiaming Cheng\\
  \textit{Independent Researcher}
  \And
  Subhransu Das\\
  \textit{The Ohio State University}
  \And
  Rajiv Ramnath\\
  \textit{The Ohio State University}
}
\date{}

\renewcommand{\shorttitle}{When Does Latent Communication Pay?}

\begin{document}

\maketitle

\begin{abstract}
Multi-agent LLM systems relay key--value caches instead of text and
credit their gains to exchanged ``latent thoughts''. That credit is a
claim about \emph{which} example's cache is relayed, not merely that
one is. We audit it causally in released systems. The cache is
replaced with deranged (mismatched-example), zeroed, and
moment-matched random counterparts, under two regimes defined by whether the receiver
needs the sender's private information. Where it does, the battery reads ceiling: 100\%
against 23--25\% for answer-irrelevant relays on the primary backbone,
a contrast replicated across three families, five checkpoints, and a
prose document-QA surface. Where it does not,
a pre-registered five-seed protocol establishes equivalence within
2.8 points, a margin anchored to the audited system's reported gain,
under Holm-corrected TOST on GSM8K and ARC-Challenge across three Qwen3
scales and on MedQA at 8B (one cell shows a small detected advantage
inside the margin); a second family shows no detected
advantage.
A large cache effect need not be a pairing effect. In one natural
cell, zeroing the relay costs 14.7 points; a mismatched cache, 0.4.
Nor is need sufficient: under the same test, delivered channels span
ceiling (LatentMAS's native relay), partial (KVComm's layer subset),
and no detected example-specific transfer (C2C's released projector).
Benchmark deltas do not by themselves establish latent-thought
transmission; establishing it takes a mismatched-cache audit, which
we release.
\end{abstract}

\section{Introduction}

When a new communication channel improves a system and the improvement
is credited to \emph{what the channel transmits}, the credit is a causal
claim the improvement alone cannot establish: the same benchmark delta
is compatible with content the receiver uses and with generic interface
effects that merely perturb the receiver in helpful ways. A direct
causal test intervenes on the transmitted content itself. Hold the
system fixed and deliver content that is equally well-formed but
belongs to the wrong example. We build that test for multi-agent LLM
systems that relay internal state between agents in place of visible
text. The relayed object is the KV cache, the per-layer attention
key--value tensors a transformer accumulates as it processes a
sequence, and recent systems attribute their reported gains to this
``latent communication'' \citep{zou2025_latent,fu2025_cache,shi2025_kvcomm}.

Auditing this attribution matters because the channel is not free:
latent relay avoids the loss and cost of re-encoding working state as
text \citep{zou2025_latent,fu2025_cache,shi2025_kvcomm}, but locks
sender and receiver to compatible checkpoints (absent a learned
cross-model projector) and leaves no directly human-readable
transcript. What justifies
attributing the gains to the relayed content, rather than to an
interface adopted for efficiency, is the claim that the cache carries
example-specific content the receiver uses. That is a causal claim
about \emph{which} example's cache is relayed, not merely that one is.
This paper audits it in delivered, released systems, without
retraining or modifying them.

Our audit intervenes on the relayed cache itself, swapping in
deranged (mismatched-example), zeroed, and moment-matched random
counterparts under two regimes defined by whether the receiver
needs the sender's private information:

\begin{itemize}
\item \textbf{Calibrated regime} (procedurally generated sender-private
facts, receiver cannot reconstruct): the instrument reads a
ceiling-level effect. Aligned relay reaches $\approx$100\% against
$\approx$23--25\% for answer-irrelevant relays on Qwen3-8B, and the
contrast replicates across three model families spanning five checkpoints:
Qwen3-4B/8B/14B, Mistral-Nemo-12B, and phi-4.
\item \textbf{Natural regime} (standard benchmarks, receiver could solve
alone): the \emph{same} intervention reads \emph{zero} to within the
margin. Holm-corrected
seed-level equivalence testing bounds the cache-matching contrast
within $\pm2.8$ points, a margin operationalized from the audited
system's reported gain, on GSM8K and ARC-Challenge across three Qwen3
scales and on MedQA at 8B (one cell shows a small detected advantage
inside the margin); a second family shows no detected advantage.
Random caches collapse accuracy, which confirms that the intervention
reaches the receiver.
\end{itemize}

What flips between the regimes is not whether communication
``helps'' but whether \emph{example pairing} carries value. A large
cache effect need not be a pairing effect:
in one natural cell zeroing the relay costs the receiver 14.7 points
while a mismatched cache makes no detectable difference. The
equivalence bound targets precisely the pairing term, the one the
``latent thoughts'' attribution needs. The consequence is a standard of
evidence: headline gains on standard benchmarks cannot, by
themselves, be read as evidence of latent-channel transmission.

The audit also ports across systems, where it dissociates a second
time. Receiver need, while decisive for the audited native relay, is
not sufficient: under the same information-asymmetry test, three
delivered channels read ceiling, partial transfer, and no detected
example-specific transfer. An instrument that was simply blind, or one that measured
only the presence of a cache prefix, could produce neither
dissociation.

\subsection{Contributions}
\begin{itemize}
\item The receiver-necessity dissociation. Across five checkpoints
from three model families and three task surfaces (math,
multiple-choice science, prose counterfactual QA), the same cache
intervention finds example-pairing value at or near ceiling where the
receiver needs the sender's private content, and bounded or
undetected where it does not.
\item A causal evidence standard, and the audit methodology that
operationalizes it. Gains credited to transmitted content require a
mismatched-content control; ours combines cache-level interventions
(derangement, zeroing, moment-matched randomization) with a procedurally
generated sender-private calibration instrument that separates
receiver redundancy from channel failure.
\item Portability of the audit to two further released
latent-communication systems, C2C and KVComm: under the same test,
the three delivered channels read three distinct levels of detected
content transfer.
\item Released calibration instrument and audit harness.
\end{itemize}

\section{Related Work}

\paragraph{Multi-agent LLM collaboration.}
Debate, ensembling, and mixture-of-agents exchange visible messages and
report accuracy gains from doing so
\citep{du2023_improving,wang2024_mixture,jiang2023_llm}, with
self-consistency as the single-model analogue \citep{wang2022_self} and
routing deciding when collaboration is worth invoking
\citep{chen2023_frugalgpt,ong2024_routellm}. Recent negative results
temper the picture: committee members can fail to explore one another's
capabilities \citep{choi2026_multi} and representations can collapse
toward redundancy \citep{patel2026_representational}. Our audit asks
the corresponding causal question for the \emph{latent} variant.

\paragraph{Latent communication in multi-agent LLM systems.}
A growing line of systems relays internal state between agents in place
of visible text: LatentMAS prepends the sender's layer-wise KV cache to
the receiver's \citep{zou2025_latent}; C2C learns a cross-model
projector into a different receiver's cache space \citep{fu2025_cache};
KVComm relays a calibration-selected layer subset \citep{shi2025_kvcomm}.
A broader family relays activations or hidden ``thoughts'', hybridizes
latent and text protocols, extends to memory and multimodal settings,
or compresses the relay
\citep{ramesh2025_communicating,du2025_enabling,zheng2025_thought,mou2026_hylat,fu2026_latentmem,liu2026_vision,chen2026_see,li2026_when};
see \citet{liu2026_beyond,chen2026_five} for surveys. The lineage runs
back to emergent communication in multi-agent RL
\citep{foerster2016_learning,sukhbaatar2016_learning,lazaridou2016_multi,mordatch2017_emergence}.
These systems share the attribution under audit here: that relayed
latent state transmits example-specific content the receiver uses. We
test that attribution in three delivered systems rather than proposing
a competing relay; Appendix~\ref{app:related} extends this survey.

\paragraph{Latent reasoning.}
Distinct from inter-agent relay, latent reasoning recycles hidden
states within a single model \citep{hao2024_training}. LatentMAS
combines both components; our audit isolates the communication
component by holding reasoning machinery fixed within every arm
contrast.

\paragraph{Cache- and activation-level interventions.}
KV-cache and activation editing has been used for efficiency and for
safety auditing of shared caches
\citep{an2026_rest,bui2026_make,asif2026_lcguard,brito2026_when,wang2026_out}.
Methodologically closest to us is activation patching / causal tracing,
which swaps internal states between runs to localize causal
contributions
\citep{meng2022_locating,wang2022_interpretability,zhang2023_towards,geiger2021_causal};
cross-model transfer of internal states has been probed directly with
both positive \citep{oozeer2025_activation,chen2025_transferring} and
negative \citep{zhang2026_negative} findings. Our derangement arm is an
inter-\emph{agent} analogue, extending the patching logic from model
internals to the communication boundary between agents while preserving
the cache distribution.

\paragraph{Equivalence testing in empirical ML.}
TOST-style equivalence protocols with pre-registered margins give nulls
quantitative content
\citep{schuirmann1987_comparison,lakens2017_equivalence},
with Holm step-down correction for multiple comparisons
\citep{holm1979_simple}. We adopt this machinery, anchoring the margin
to the audited system's own claimed gain.

\paragraph{Relation to the system's own ablation.}
LatentMAS's hybrid ablation (v3, appendix Table~7) compares
communication \emph{media}; our audit intervenes on cache
\emph{content} within the delivered latent interface, so both results
can hold simultaneously (comparison detail in
Appendix~\ref{app:hybrid-ablation}).

\paragraph{Concurrent work.}
Concurrent with this work, \citet{zhang2026_do} audit LatentMAS's
relay on Qwen3-4B/8B (GSM8K, ARC-Challenge, MATH-500) with controlled
message replacements whose other-example message parallels our
deranged arm, and reach the same motivating conclusion: aggregate
accuracy does not identify what a latent message contributes. The
audits differ in scope and inference. Ours spans three released
systems and five checkpoints from three families, manipulates
receiver necessity directly, and adds zeroed and moment-matched
random arms as primary controls; and where their example-level
contrasts (at most three decoding seeds per cell, one on MATH-500)
leave practical equivalence unestablished at a declared
$\pm1$-point margin, our pre-registered protocol of five seeds at
$n = 500$ per arm per seed, with seed-level inference, bounds the
pairing term in the Qwen3 cells below the $\pm2.8$-point gain the
audited system itself claims.

\section{Audit Design}

\begin{figure}[t]
\centering
\includegraphics[width=\textwidth]{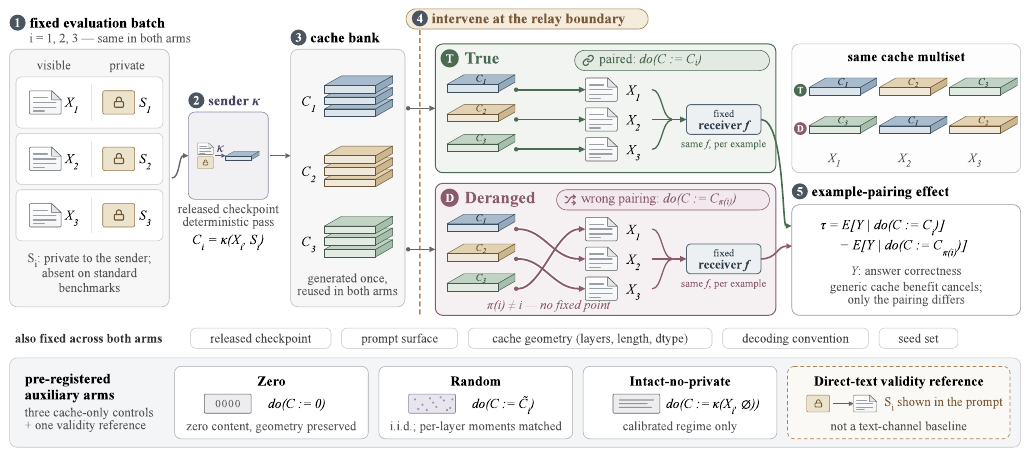}
\caption{\textbf{Causal audit design.} The primary contrast holds the
receiver and the batch of caches fixed, permuting only which cache
reaches each query. Both arms relay an equally well-formed cache, so
the true-minus-deranged difference isolates example pairing from any
generic benefit of relaying one.}
\label{fig:audit}
\end{figure}

\paragraph{Audited mechanism.}
In the delivered LatentMAS pipeline, agents are instances of the same
released checkpoint arranged sequentially; after an upstream (sender)
agent processes its context, its layer-wise key--value cache is extracted
and prepended, layer by layer, to the downstream (receiver) agent's cache
before the receiver generates. The system's stated position is not a
gloss but an information-preservation theorem: generating from the
relayed cache is equivalent to receiving the sender's outputs directly,
``latent thoughts'' included \citep[Theorem~3.3]{zou2025_latent}, and
downstream gains are credited to that content.
Our audit intervenes exactly at the relay boundary
(Figure~\ref{fig:audit}). Every arm delivers a
cache with identical geometry, meaning the same layer count, sequence
length, and dtype, to the same receiver under the same prompt surface.
Any accuracy difference between arms is therefore attributable to cache
\emph{content}.

\paragraph{Intervention arms.}
The full taxonomy comprises six arms; each cell pre-registers the subset
it uses. \textbf{True}: the delivered cache for the paired example.
\textbf{Deranged}: the delivered cache of a \emph{different} example from
the same batch (a constrained permutation with no fixed point). This
is the canonical content control, since it preserves the marginal
cache distribution exactly. \textbf{Zero}: all cache entries zeroed.
\textbf{Random}: Gaussian cache entries centered and rescaled per
example, layer, and K/V tensor to match the true cache's moments. \textbf{Intact-no-private}: the sender
processes the example with the sender-private information replaced by
a matched answer-irrelevant counterpart, in
the calibrated regime only, which isolates answer-relevant content
from generic sender computation. \textbf{Direct-text reference}: the sender-private
information is placed verbatim in the receiver's visible prompt. This
last arm is a validity calibration, not a medium comparison: it changes
both information availability and the terminal prompt surface, and we
never interpret it as a text-channel baseline.

\paragraph{The calibration instrument.}
On standard benchmarks a cache-matching contrast is uninterpretable
alone: a null is equally consistent with receiver redundancy and with
channel failure. We therefore construct a regime in which the receiver
\emph{cannot} solve the task without the relay. Each example instantiates a procedurally generated registry of
entity--attribute bindings, sampled fresh at evaluation time, so the
sampled bindings are absent from any training corpus except with
vanishing probability; the sender observes the
registry, the receiver observes only the query. The measured gap
between the true arm and the content controls is then a direct reading
of how much sender-private, answer-relevant content the channel
delivers. Concretely, a registry
is four bindings of the form ``\emph{$\langle$entity$\rangle$ keeps
signal word $\langle$value$\rangle$}'', with both sides seven-letter
strings sampled fresh from a fixed alphabet (e.g., \texttt{invyqqz}
$\to$ \texttt{zevrlel}); the query names one value and asks which of
the four entities, relabeled A--D, holds it. The intact-no-private
control relays a registry with identical format whose bindings
cannot answer the query.

\paragraph{Estimands.}
Fix a cell (backbone, task, regime) and its frozen seed set. For
example $i$, let $X_i$ be the receiver-visible surface (prompt and
query), $S_i$ the sender-private information (the registry in the
calibrated regime; absent on standard benchmarks), and
$C_i = \kappa(X_i, S_i)$ the delivered sender cache (the sender pass
is deterministic). Let $A_i$ be the correct answer and
$\hat{A} = f(X_i, C, \varepsilon)$ the fixed receiver's prediction
under whatever cache $C$ the arm delivers, with $\varepsilon$
collecting decoding randomness; correctness is
$Y = \mathbf{1}\{\hat{A} = A_i\}$. Every audit arm except the
direct-text validity reference intervenes on $C$ alone:
$do(C{:=}C_i)$ (true), $do(C{:=}C_{\pi(i)})$ (deranged),
$do(C{:=}\mathbf{0})$ (zero), $do(C{:=}\tilde{C}_i)$ (random), and, in
the calibrated regime, $do(C{:=}\kappa(X_i, S_i^0))$
(intact-no-private), where $S_i^0$ is the matched answer-irrelevant
registry, holding $f$, $X_i$, and cache geometry fixed.
The primary estimand is the \emph{pairing effect}
\begin{equation}
\tau \;=\; \mathbb{E}\big[Y \mid do(C{:=}C_i)\big]
\;-\; \mathbb{E}\big[Y \mid do(C{:=}C_{\pi(i)})\big],
\label{eq:pairing}
\end{equation}
with expectations over examples and decoding randomness within a seed.
Writing $\tau_{AB}$ for the contrast between arms $A$ and $B$, the arm
contrasts satisfy the accounting identity
\begin{equation}
\underbrace{\tau_{TZ}}_{\text{true vs.\ zero}}
\;=\;
\underbrace{\tau_{TI}}_{\text{true vs.\ no-private}}
\;+\;
\underbrace{\tau_{IZ}}_{\text{no-private vs.\ zero}}
\label{eq:decomp}
\end{equation}
in the calibrated regime, and $\tau_{TZ} = \tau_{TD} + \tau_{DZ}$ in
the natural regime. The identities are algebraic, not an assumption of
additive mechanisms; what makes their terms meaningful is the arm
construction. Intact-no-private differs from the true arm only in
replacing $S_i$ with the answer-irrelevant $S_i^0$ in the sender's
context, and from the zero arm only
in carrying a well-formed sender computation; $\tau_{TI}$ therefore
reads as answer-relevant-content value and $\tau_{IZ}$ as
generic-computation value.
Only the content terms ($\tau_{TI}$, $\tau_{TD}$) are the ``latent
thoughts'' attribution under audit; the generic terms measure the
value of receiving \emph{some} well-formed sender computation.

\paragraph{Identification.}
Two observations carry the identification. \emph{First, the deranged
arm is marginal-preserving:} $\pi$ is a fixed-point-free permutation of
the batch, so the deranged arm delivers exactly the true arm's multiset
of caches and only the pairing of cache to query changes. A nonzero
$\tau$ therefore cannot be produced by cache malformedness or
distribution shift; it isolates the effect of correct pairing.
\emph{Second, the regimes assign $\tau$ complementary burdens of
proof.} In the calibrated regime, $S_i$ is sampled fresh at evaluation
time, is unpredictable from $X_i$, and reaches the receiver only
through the relay. Under the control arms, zero and intact-no-private,
the delivered cache is not a function of $S_i$, so the receiver's
\emph{prediction} satisfies $\hat{A} \perp S_i \mid X_i$, while
correctness still depends on $S_i$ through the gold answer. That is
exactly why control-arm accuracy on $S_i$-determined queries falls to
the level achievable without the private content. True-arm accuracy
far above that level shows the channel delivers answer-relevant
private content: the instrument is demonstrably not blind in this
regime. In the natural regime the attribution predicts $\tau > 0$, and
a frozen TOST that rejects both one-sided nulls replaces a bare
nonrejection with the affirmative bound $|\tau| < \delta$; the random
arm, whose cache retains only per-layer moments, collapses accuracy
and shows the intervention point is causally coupled. A
natural-regime null is therefore a calibrated zero rather than a dead
instrument. Jointly the regimes rule out the global form of both
failure modes of a single-regime audit: an instrument insensitive by
construction would read zero in the calibrated regime, and one that
manufactures signal would read nonzero in the natural regime.
Regime-specific insensitivity is not excluded by the dissociation
alone; it is bounded separately by the TOST interval width in each
natural cell. The audit's headline pattern is then a statement about
Eq.~\ref{eq:decomp}: the content term is at or near ceiling in every
calibrated cell, bounded below the operationalized margin in the
Qwen3 natural cells, and undetected in the second family's cells,
where no bound is established. The generic term meanwhile varies
freely. On MedQA the total effect sits almost entirely in the
non-pairing terms: true versus zero is $+14.7$ points, against
$\tau_{TD} = +0.4$ with 90\% CI $[-1.34, +2.14]$, both seed-level
means from the raw records.

\paragraph{Statistical conventions.}
All cells share frozen conventions, fixed before any production run was
unblinded: unconditional accuracy as the primary outcome with parse
failures scored wrong; $n{=}500$ examples per arm per seed in a fixed
$25{\times}20$ batch geometry; three receiver-sampling seeds (42/43/44)
per cell, expanded to five seeds only in the pre-registered natural-audit
equivalence family; receiver decoding at the audited system's released
sampling configuration (temperature $0.6$, top-$p$ $0.95$), the sole
seed-varying randomness; Holm correction within each pre-registered
hypothesis family; and no post-hoc arm acceptance. Equivalence claims use two
one-sided tests (TOST) with margin $\pm 2.8$ percentage points, anchored
to the audited system's own claimed sequential gain, so that a passing
test bounds the relay's example-pairing contribution below the gain the
system attributes to it. Concretely, LatentMAS reports an aggregate
sequential gain of 2.8\% over its text-based counterpart across its
nine-benchmark suite \citep[\S4.1]{zou2025_latent}; we operationalize
that reported figure as an equivalence margin of $\pm2.8$ percentage
points on the accuracy scale, the scale on which the system's
per-task improvements are tabulated. The margin was frozen before
any unblinding. We anchor to the aggregate rather than to per-task
figures: the tabulated per-task improvements are variance-free point
estimates, and on the audited Qwen3 task--backbone cells they run
from $-1.6$ to $+2.3$ points, several negative --- a negative claimed
gain yields no margin at all. No audited cell reports a per-task gain
outside the margin.
The system's gains are test-split figures; the GSM8K and
ARC-Challenge cells use fixed training-split items (MedQA: test
split), so the bound reads on the audited item sets. Any memorization
of training items raises receiver redundancy, which works toward the
null the bound reports, not against it.
All reported numbers are recomputed from raw per-example records; a
frozen replacement rule governs incomplete grids (a seed whose arm
grid cannot complete is replaced wholesale before unblinding), and
batch geometry is never reduced to fit memory. Per-cell GPU venues,
the pinned software stack, and the full run history are disclosed in
Appendix~\ref{app:reproducibility}.

\section{Native Relay: Receiver-Necessity Dissociation}

\subsection{Calibrated Regime: The Instrument Reads Ceiling}

When the receiver needs the relayed content, the intervention battery
reads an at-or-near-ceiling effect (Figure~\ref{fig:dissociation},
upper block), and does so across three model families
spanning five checkpoints: Qwen3-4B/8B/14B
\citep{yang2025_qwen3}, Mistral-Nemo-12B \citep{mistral2024_nemo},
and phi-4
\citep{abdin2024_phi}. The Qwen3 scale span rules out a size artifact;
the two further families, a family artifact.

\paragraph{Primary cell (Qwen3-8B).}
The delivered aligned relay reaches 500/500 on all three seeds; the
matched answer-irrelevant relay (intact-no-private) reaches 126/122/117
(23.4--25.2\%). The paired true-minus-intact difference is
$+74.8/{+}75.6/{+}76.6$ points per seed.

\paragraph{Family replication.}
The contrast replicates on every family tested
(Table~\ref{tab:calibrated}): true-minus-intact differences of $+75.6$
to $+78.8$ points on Qwen3-4B and Qwen3-14B, and $+58.2$ to $+63.4$ on
Mistral-Nemo-12B, the smallest but still decisive margin --- and the
one checkpoint whose true arm sits below ceiling ($84$--$86\%$ against
its direct-text arm at $\geq 499/500$), so relay fidelity rather than
receiver need is what binds there. Two
quantities do not appear in the table: 14B's simultaneous lower bound
of $0.68$ on the true-minus-intact contrast, including worst-case
parse resolution, and phi-4's true-minus-zero contrast of
$\approx{+}1.00$, the strongest ceiling observed.

\begin{table}[t]
\centering
\small
\setlength{\tabcolsep}{5pt}
\begin{tabular}{lcccccrrr}
\toprule
Family & True & Direct & Intact & Deranged & Zero &
T$-$D & T$-$I & I$-$Z \\
\midrule
Qwen3-4B         & 498/499/498 & 494/494/492 & 116/105/120 & 42/55/37    & 20/17/18 & $+90.7$ & $+76.9$ & $+19.1$ \\
Qwen3-8B         & 500/500/500 & 500/500/500 & 126/122/117 & 53/60/53    & 8/11/8   & $+88.9$ & $+75.7$ & $+22.5$ \\
Qwen3-14B        & 500/500/499 & 500/500/500 & 110/114/117 & 62/68/75    & 47/43/50 & $+86.3$ & $+77.2$ & $+13.4$ \\
Nemo-12B         & 425/431/420 & 500/499/500 & 108/125/129 & 115/124/123 & 0/0/0    & $+60.9$ & $+60.9$ & $+24.1$ \\
phi-4            & 499/500/500 & 500/500/500 & 110/125/118 & 64/63/67    & 0/0/0    & $+87.0$ & $+76.4$ & $+23.5$ \\
\bottomrule
\end{tabular}
\caption{Calibrated (receiver-necessity) cells: correct answers out of
500 per arm, seeds 42/43/44, with the seed-mean contrasts of
Eq.~\ref{eq:decomp} in percentage points (T$-$D = true$-$deranged
pairing contrast; T$-$I = true$-$intact content term; I$-$Z =
intact$-$zero generic term). True = delivered aligned relay; Direct =
direct-text validity arm; Intact = matched answer-irrelevant relay;
Deranged = content from a mismatched registry example; Zero =
geometry-preserving zeroed relay. Nemo-12B is Mistral-Nemo-12B.}
\label{tab:calibrated}
\end{table}

\paragraph{The receiver reads the channel.}
Ceiling accuracy shows the channel delivers content but not yet that
the receiver \emph{reads} it. Under deranged relays, receiver outputs
shift toward the \emph{donor} example's registry rather than the
queried one: across seeds, $64.2$--$64.8\%$ of outputs match the
donor's label versus $20.8$--$23.4\%$ under the matched
answer-irrelevant control. The receiver consumes relayed content even
when that content misleads.

\paragraph{Why the zero and intact arms matter.}
Across families the zero arm ranges from 0/500 to $\approx$47/500 and
intact-no-private sits well above zero but far below true. The
ordering true $\gg$ intact $>$ zero splits the relay's effect along
Eq.~\ref{eq:decomp} into generic sender-computation value and
sender-private content value, and the latter is the dominant term in
every family.

\paragraph{Surface-form stress test: the ceiling survives a prose
surface.}
A skeptical reading of the calibrated ceiling is that it is an artifact
of the registry grammar, a lookup-table surface. To test surface-form
generality we rebuilt the calibration as counterfactual document QA:
procedurally generated, fixed-frame, four-sentence prose paragraphs
embedding the sender-private bindings, with position-matched tokenizer
geometry. We ran the same five-arm battery on Qwen3-8B (three seeds,
$n{=}500$ per arm per seed). The intervention swaps the donor's
complete post-refiner relay state (all layers, keys and values, the
attention-mask row, and the next semantic position) at the
refiner-to-judger boundary,
immediately before the judger generation call; the judger's prompt,
query, and scoring are unchanged.
The ceiling transfers intact: the true relay and the direct-text
reference read 500/500 on every seed, while the intact-no-private arm
falls to 22/31/20 correct, the deranged arm to 31/31/26, and the zero
arm to 30/23/33. Simultaneous paired-advantage lower bounds exceed
$0.90$ against every control arm (true-vs-intact $0.918$,
true-vs-deranged $0.912$, true-vs-zero $0.907$; three-seed
intersection--union test, which also passes under the worst-case parse
convention). Every control arm sits far below the pre-registered 31\%
interpretation cutoff, which bounds the strongest audited uninformed
heuristic on the construction ($28.6\%$), frozen before the audit ran. What transfers here is surface form, from
registry lines to grammatical paragraphs under the same constructed
information asymmetry. This is not a synthetic-to-natural transfer, and
we make no claims about human-authored documents.
Option-only elimination is common across the matched arms because the
public query and choices are unchanged. Relay-conditioned elimination
may differ by arm and is part of the intervention; the deranged relay
may actively mislead and is not treated as a no-information floor. The
matched no-private relay may also be behaviorally nonneutral, so the
zero-cache control is reported separately. The cell's finalizer re-parses every raw
output under the frozen grammar and recomputes correctness from
canonical gold records; it establishes self-consistency of the
attested outputs, not generation fidelity.

\subsection{Natural Regime: Bounding the Pairing Term}

\begin{figure}[t]
\centering
\includegraphics[scale=1]{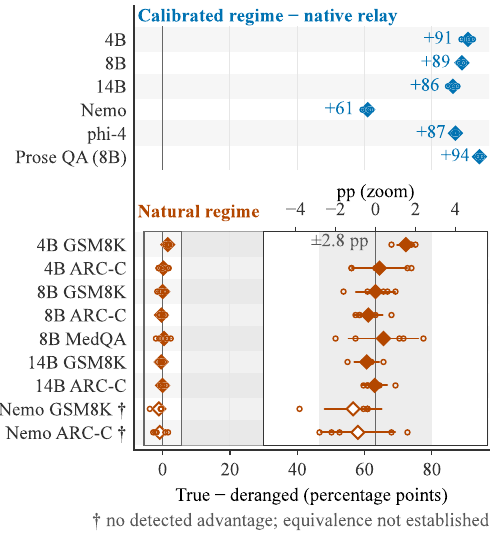}
\caption{\textbf{One estimand, two regimes.} True-minus-deranged seed
means (diamonds; per-seed contrasts as circles) span $+61$ to $+94$
points where the receiver needs sender-private content and cluster
near zero where it solves alone. The zoomed panel magnifies the
natural block, showing unadjusted 90\% $t$ CIs against the
$\pm2.8$-point margin operationalized from the audited system's
reported gain; hollow diamonds mark cells where Holm-corrected TOST
does not establish equivalence.}
\label{fig:dissociation}
\end{figure}

On GSM8K \citep{cobbe2021_training} and
ARC-Challenge \citep{clark2018_think} (fixed training-split items),
the receiver can solve the task without the sender. The instrument that just read ceiling now
reads zero to within the operationalized margin, with one small
detected advantage inside it
(Figure~\ref{fig:dissociation}; Table~\ref{tab:natural}). If the relayed cache transmits
example-specific content the receiver uses, replacing it with a
deranged counterpart should cost accuracy, the pairing effect of
Eq.~\ref{eq:pairing}; the equivalence protocol of
the Audit Design section asks
whether that cost is bounded away from the audited system's own claimed
gain.

\begin{table}[t]
\centering
\small
\setlength{\tabcolsep}{5pt}
\begin{tabular}{llrrrcrrrl}
\toprule
Cell & Task & True & Der. & T$-$D & 90\% CI & Raw $p$ & Holm $p$ & 2-sided $p$ & Verdict \\
\midrule
Qwen3-4B  & GSM8K & 90.76 & 89.24 & $+1.52$ & $[+1.08, +1.96]$ & .0017 & .0034 & .0018 & eq.\ + adv. \\
Qwen3-4B  & ARC-C & 88.28 & 88.08 & $+0.20$ & $[-1.19, +1.59]$ & .0081 & .0081 & .774 & eq. \\
Qwen3-8B  & GSM8K & 93.68 & 93.68 & $0.00$  & $[-0.98, +0.98]$ & .0018 & .0036 & 1.00 & eq. \\
Qwen3-8B  & ARC-C & 96.60 & 96.96 & $-0.36$ & $[-1.08, +0.36]$ & .0010 & .0030 & .346 & eq. \\
Qwen3-8B  & MedQA & 75.20 & 74.80 & $+0.40$ & $[-1.34, +2.14]$ & .0213 & .0213 & .650 & eq. \\
Qwen3-14B & GSM8K & 93.56 & 94.00 & $-0.44$ & $[-1.06, +0.18]$ & .0006 & .0007 & .207 & eq. \\
Qwen3-14B & ARC-C & 96.40 & 96.44 & $-0.04$ & $[-0.66, +0.58]$ & .0004 & .0007 & .898 & eq. \\
Nemo-12B  & GSM8K & 79.20 & 80.32 & $-1.12$ & $[-2.55, +0.31]$ & .0333 & .0665 & .170 & not est. \\
Nemo-12B  & ARC-C & 13.44 & 14.32 & $-0.88$ & $[-2.74, +0.98]$ & .0464 & .0665 & .370 & not est. \\
\bottomrule
\end{tabular}
\caption{Natural-regime cells, full statistics (five seeds each):
true and deranged arm mean accuracies (\%), seed-mean
true$-$deranged (points), unadjusted 90\% $t$ CI, raw and
Holm-adjusted TOST $p$ at the $\pm2.8$-point margin, two-sided
paired $t$ $p$, and the licensed verdict (eq.\ = equivalence within
$\pm2.8$; + adv.\ = with detected small advantage; not est.\ =
equivalence not established). Nemo-12B ARC-Challenge absolute levels
are interpretability-limited (Appendix~\ref{app:seed-ledger}).
Per-seed accuracies in Table~\ref{tab:seed-td}.}
\label{tab:natural}
\end{table}

\paragraph{Qwen3 family (five seeds per cell).}
All seven Qwen3 cells establish equivalence within $\pm2.8$ points
(Table~\ref{tab:natural}; Holm within each backbone's pre-registered
family, exact over the primary backbone's three tasks, where raw
TOST $p=.0018$ and $.0010$ on GSM8K and ARC-Challenge).\footnote{One 8B seed was replaced wholesale before unblinding
under the frozen replacement rule (Statistical conventions); full
run record in Appendix~\ref{app:reproducibility}.} On Qwen3-8B the two-sided
test rejects on neither task and seed signs are mixed, so no
directional wording is licensed: any latent-relay advantage on these
benchmarks is statistically bounded below the system's own
operationalized claimed gain. MedQA-USMLE (4-option English test
split) \citep{jin2020_what} closes the 8B family (same four-arm
battery, seeds 42--46): the two-sided test does not reject
($p=.65$) and seed signs are mixed: pure equivalence, no directional
wording. The random arm collapses to $0/2500$ across seeds, and,
unlike GSM8K and ARC, the zero arm costs the receiver visibly
(true $\approx 74$--$76\%$ versus zero $\approx 59$--$63\%$), a
secondary contrast showing the battery remains live on this surface
while the content-specific reading stays null. The one directional
reading sits on Qwen3-4B GSM8K, where the two-sided test rejects
with every seed positive and the pre-registered rule licenses the
compound reading: equivalent within $\pm2.8$ points \emph{with} a
detected small advantage of $+1.52$ points, about half the
operationalized margin; on 4B ARC-Challenge and
both Qwen3-14B tasks no directional wording is licensed. Across the
three Qwen3 scales, equivalence holds with at most one small bounded
advantage.

\paragraph{Mistral-Nemo-12B (five seeds).}
The same protocol on a second family reads no detected advantage
(two-sided $p=.17/.37$), with equivalence within $\pm2.8$ points
\emph{not established}: raw TOST $p<.05$ on both tasks with CIs
inside the margin, but the Holm-adjusted $p$ misses the frozen
threshold (Table~\ref{tab:natural}). We report this cell exactly as
the pre-registered rule
scores it rather than as a near miss.\footnote{Nemo's ARC absolute
accuracies fall below four-choice chance in all arms because
parse failures are scored wrong under the frozen convention; absolute
levels in that cell are interpretability-limited, while within-cell
contrasts remain valid, since all arms share the parser.}
Secondary contrasts show the instrument is live in the same cell:
true-minus-zero $+78.1/{+}13.1$ points and true-minus-random
$+77.9/{+}13.4$ (GSM8K/ARC). Both corruptions land generation at the
same $\approx\!1\%$ level (Table~\ref{tab:seed-zr}), so the two
contrasts read one collapse against two corruptions.

\paragraph{Random caches collapse accuracy.}
In every natural cell the moment-matched random arm collapses (8B:
$\leq 2/500$ per seed on both tasks; 4B: $\leq 9/500$ on GSM8K, $0/500$
on ARC; 14B: $\leq 1/500$ on GSM8K, $0/500$ on ARC; per-seed values
in Table~\ref{tab:seed-zr}). The intervention
reaches the receiver's computation, so the natural-regime null is a
calibrated zero: the channel is causally coupled, and what fails to
appear is specifically the marginal value of \emph{which} example's
cache was relayed.

\section{Porting the Audit: Three Delivered Channels}

\begin{figure}[t]
\centering
\includegraphics[scale=1]{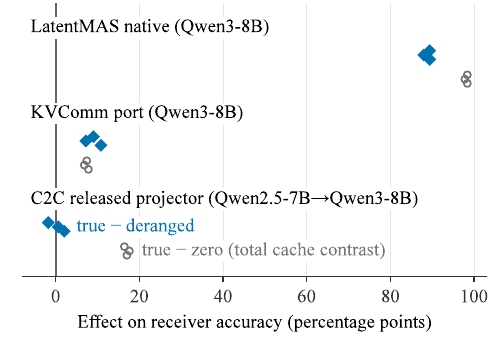}
\caption{\textbf{One asymmetry test, three delivered channels.}
Per-seed true-minus-deranged (diamonds; the pairing estimand) and
true-minus-zero (circles; total cache contrast) under the same
receiver-necessary construction, with direct-text validity arms at
ceiling throughout. C2C's sign-mixed pairing contrasts beside its
large total cache contrast read as no detected example-specific
transfer.}
\label{fig:delivered}
\end{figure}

\subsection{C2C: A Learned Cross-Model Projector}

C2C replaces the same-checkpoint KV prepend with a learned cross-model
projector \citep{fu2025_cache}. We used the released configuration,
Qwen2.5-7B-Instruct $\to$ Qwen3-8B, with the released projector weights
and fail-closed integration flags, and ran the full calibrated battery
(five arms, three seeds, $n{=}500$ per arm per seed).

Here the instrument reads no detected transfer of sender-private
content (Figure~\ref{fig:delivered}). Task and parser health check out, with the direct-text
validity arm at 499/498/500 out of 500 across seeds, but the cache arms
cluster together: true 114/114/117, deranged 104/111/126,
intact-no-private 99/108/113. True-minus-deranged comes to
$+2.0/+0.6/-1.8$ points across seeds, mixed in sign. All cache arms sit
well above the zero arm at 29/26/35, or $\approx$6\%, a large total
cache contrast --- though the cache arms themselves sit at the level
this audit's own answer-irrelevant controls reach on the same
receiver, so the contrast separates a well-formed cache from a
degenerate one, not content from its absence. What we cannot detect
is any arm separation attributable to \emph{which example's} content
was projected.

One instrument, one registry construction, one receiver family: it
reads $\approx{+}75$-point content effects through LatentMAS's KV
prepend against a sign-mixed null through C2C's learned projector in
its released configuration. We scope this precisely: it is a statement
about sender-private, answer-relevant content under our registry
construction, not a claim that C2C's reported benchmark gains are
artifactual.

\subsection{KVComm: Layer-Selected KV Relay}

KVComm relays only a calibrated subset of layers: a selection score
combining attention importance on a calibration set with a depth prior
picks the top 30\% of layers \citep{shi2025_kvcomm}; the delivered
code also always transfers layer 0, for 11 effective layers of 36
here.
We evaluate a Qwen3-8B port of the delivered KVComm selection and
transfer algorithms; the one change that could plausibly move the
number is replacing evaluation-prefix calibration with a
pre-specified disjoint 100-example split (port mechanics in
Appendix~\ref{app:kvcomm-port}). Layer selection was calibrated once under
the true-private condition on a disjoint sample from the same
receiver-necessity generator; all reported effects are therefore
conditional on this task-matched selector and do not establish
task-agnostic or cross-task layer selection.

Here the instrument reads a third, intermediate pattern: detected
transfer, far below ceiling. Under the full calibrated battery (five
arms, three seeds, $n{=}500$) the direct-text validity arm is at
499/500 in every seed, and the true-private relay reaches 165/153/156, against
111/117/111 for deranged, 127/124/106 for intact-no-private, and
126/119/119 for zero. Every content contrast is positive in every seed
(true$-$deranged $+10.8/{+}7.2/{+}9.0$ points; true$-$intact
$+7.6/{+}5.8/{+}10.0$; true$-$zero $+7.8/{+}6.8/{+}7.4$), so the
11-layer relay demonstrably carries example-specific content; but where
LatentMAS's full-cache prepend converts the same construction into
$\approx{+}75$-point effects at ceiling, the layer-selected relay
recovers roughly a tenth of that. Notably, the zero arm sits level with
the other content controls rather than below them: unlike C2C, the
total cache contrast is no larger than the content contrasts, so there
is nothing content-independent to subtract. The zero arm here is a
geometry-preserving zero-content ablation, not removal of the cache
interface; the retained geometry is specified in
Appendix~\ref{app:kvcomm-port}. This KVComm cell tests relay value in a
receiver-necessary setting; because KVComm is not crossed with a matched
receiver-sufficient condition, the cell does not by itself identify a
within-KVComm necessity-by-relay interaction. Integration costs and
per-cell execution details for both ported systems are disclosed in
Appendix~\ref{app:reproducibility}.

\section{Discussion}

\paragraph{Receiver necessity moderates, and does not create, relay value.}
For the audited native relay, the instrument's reading flips with
whether the receiver needs the sender's private
information. The cross-system cells show receiver need
is not sufficient. Under the same information asymmetry,
delivered channels read ceiling, partial, and none detected, so necessity
moderates the value of a channel that demonstrably transfers content;
necessity does not create one. A pre-registered bolt-on cell
(Appendix~\ref{app:bolton}) makes the same point mechanically: prepending
latent vectors is not sufficient for example-specific transfer and can
disrupt the receiver regardless of content.
The evidence spans one relay construction family, so moderation should
be tested prospectively on independent constructions. The ``obvious'' null, that
of course the relay does not matter where the receiver solves alone, is
itself the attribution under audit, and because the margin is anchored
to the audited system's own claimed gain, the null is a quantitative
bound on it, not a rhetorical shrug.

\paragraph{Scope.}
The audit bounds the example-pairing value of delivered relays; it
issues no latent-versus-text verdict, and the calibrated cells rule
out the reading that relayed caches carry nothing. Separately,
Llama-3.2-3B is a portability boundary, not an audit result: in both
delivered realignment configurations the receiver could not consume
the relay, so the audit's precondition of a functioning channel was
not met (evidence in Appendix~\ref{app:llama}).

\paragraph{Limitations.}
Three are material. (1) External systems are audited in one released
configuration each; retrained or retuned variants could behave
differently. (2) The calibrated regime uses one registry construction;
other constructions of sender-private content could modulate the
ceiling (the prose cell partially addresses this). (3)
Absolute accuracy levels in some cells are interpretability-limited
(see the Nemo ARC footnote) and should not be compared across papers;
within-cell contrasts remain valid because the convention is applied
uniformly across arms.

\paragraph{Reproducibility.}
We release the audit harness, the calibration-instrument generator, and
per-cell run manifests with per-arm raw outcome ledgers; the full
release contents, venue table, and run-deviation record are in
Appendix~\ref{app:reproducibility}.

\section{Conclusion}

When does latent communication pay? For the audited native relay:
when \emph{which} example's cache is relayed carries value the
receiver cannot supply itself. One battery, run unchanged across two
regimes, reads both poles: at or near ceiling where the sender holds
private, answer-relevant content, replicated across three families
and five checkpoints; pairing value bounded
below the audited system's claimed gain where the receiver solves
alone, with no detected advantage in a second family. Need alone is
only half the answer; under the same test, delivered channels span
ceiling, partial, and none detected. The released instrument turns
``does latent communication work?'' from a benchmark-delta debate into
a reusable measurement. For builders: prioritize latent bandwidth
where information asymmetry is real, and demand a mismatched-cache
control as the evidence that a channel carries content.

\section*{Acknowledgments}

This work made use of computational resources at the Ohio
Supercomputer Center \citep{osc1987}, including the Ascend cluster
\citep{osc_ascend2022}. Per-cell venue assignments are in
Table~\ref{tab:venues}.

\bibliography{aaai2027}

\appendix

\section{Robustness of the Natural-Regime Bound}
\label{app:robustness}

\begin{figure}[t]
\centering
\includegraphics[scale=1]{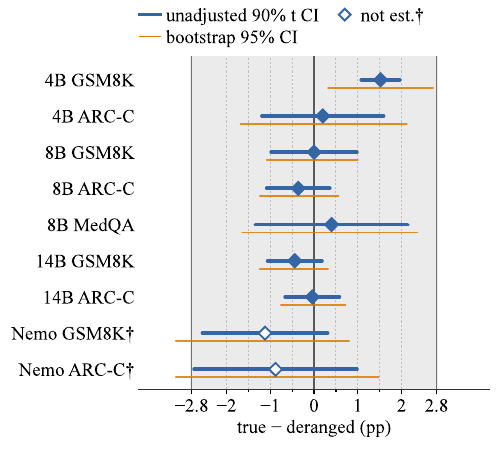}
\caption{\textbf{Margin sensitivity ladder.} All nine unadjusted
intervals lie inside the operationalized $\pm2.8$ band, two already
inside $\pm1$; the seven Qwen3 cells establish Holm-corrected
equivalence there, and the two Mistral-Nemo-12B cells do not.
Unadjusted seed-level 90\% $t$ CIs (thick bars; diamonds at the
seed-mean) for all nine natural cells (4B/8B/14B: Qwen3; Nemo:
Mistral-Nemo-12B) against the tightening margin verticals $\pm0.5$,
$\pm1$, $\pm1.5$, $\pm2$ inside the $\pm2.8$ band (shaded). Of the
seven Qwen3 cells, none fits within $\pm0.5$, two within $\pm1$, four
within $\pm1.5$, six within $\pm2$, and all seven within $\pm2.8$.
Hierarchical bootstrap 95\% intervals (thin
bars) are descriptive companions. Hollow diamonds ($\dagger$):
equivalence not established under the pre-registered Holm-corrected
readout, which adjusts what the unadjusted bars do not; Nemo-12B
ARC-Challenge absolute levels are interpretability-limited
(Table~\ref{tab:seed-td}). The ladder is descriptive; the licensed
verdicts are the Holm-corrected TOST at $\pm2.8$ only.}
\label{fig:margin-ladder}
\end{figure}

\paragraph{Margin sensitivity.}
The seven Qwen3 intervals sit well inside the margin they were
tested at: four fit within $\pm1.5$ points and six within $\pm2$.
For the raw TOST at $\alpha=.05$, a cell passes at margin $m$
exactly when its unadjusted $90\%$ interval lies within $\pm m$, so
the smallest passing margin is the interval's larger absolute
endpoint (Table~\ref{tab:natural}): $0.66$ (14B ARC-Challenge),
$0.98$ (8B GSM8K), $1.06$ (14B GSM8K), $1.08$ (8B ARC-Challenge),
$1.59$ (4B ARC-Challenge), $1.96$ (4B GSM8K), and $2.14$ (8B
MedQA). Read as a ladder over $\pm0.5$, $\pm1$, $\pm1.5$, $\pm2$,
and $\pm2.8$ points: no cell passes at $\pm0.5$; two of the seven
Qwen3 cells pass at $\pm1$; four pass at $\pm1.5$; six pass at
$\pm2$; all seven pass at the operationalized $\pm2.8$. The ladder
is descriptive; the licensed verdicts remain the pre-registered
Holm-corrected TOST at $\pm2.8$. The largest per-task gain the
system tabulates on any audited cell is $+2.3$ points (4B
ARC-Challenge), and the largest interval endpoint across the seven
Qwen3 cells is $2.14$, so all seven cells also fit within a margin
set at that cell-level anchor. The primary backbone's GSM8K and
ARC-Challenge intervals thus fit within margins roughly a third of
the operationalized one (Figure~\ref{fig:margin-ladder}).

\paragraph{Margin anchor: the aggregate gain.}
A cell-level alternative --- one margin per task from the system's
own per-task figures --- fails on those figures' properties, not on
principle. An equivalence margin is a smallest effect of scientific
interest, and the system's per-task improvements cannot supply one:
they are variance-free point estimates, they are negative on some
audited surfaces (a negative margin is undefined), and at their
smallest they are noise-scale numbers the system itself never
advances as stand-alone gains. In full, the audited cells'
tabulated per-task figures are $-1.6$ (GSM8K) and $+2.3$
(ARC-Challenge) at 4B; $+1.5$ (GSM8K), $-0.2$ (ARC-Challenge), and
$+0.3$ (MedQA) at 8B; and $+1.4$ (GSM8K) and $-0.3$
(ARC-Challenge) at 14B. Reporting the pairing effect as a
proportion of the per-task delta inherits the same defect with a
near-zero or negative denominator. The aggregate $2.8$ is the one
gain figure the system claims for the mechanism as such; the margin
tests the attribution at the level the attribution is made.

On one cell the audit's interval excludes the system's tabulated
per-task gain; on another it is compatible with it. On Qwen3-8B
GSM8K the pairing CI of $[-0.98, +0.98]$ excludes the system's
tabulated $+1.5$-point per-task improvement (an exclusion of a
variance-free point figure, reported as an observation, not a
licensed test). Conversely, on MedQA the interval $[-1.34, +2.14]$
is compatible with a pairing contribution the size of the system's
$+0.3$ figure; the licensed claim there is nothing stronger than
equivalence within the operationalized margin. The system's largest
per-task claims (e.g.\ MBPP+ $+5.1$ at 8B) sit on surfaces whose
test sets cannot support the frozen $n{=}500$ geometry (MBPP+
offers 378 problems and HumanEval+ 164); see
Appendix~\ref{app:task-selection}.

\paragraph{Multiplicity sensitivity.}
The equivalence verdicts do not depend on partitioning the Qwen3
hypotheses into per-backbone families: pooling all seven cells into
one Holm family leaves every adjusted TOST $p$-value below $.05$.
Recomputed from the per-seed ledger (Table~\ref{tab:seed-td}), the
seven-cell Holm-adjusted $p$-values are $.0025$ (14B
ARC-Challenge), $.0039$ (14B GSM8K), $.0048$ (8B ARC-Challenge),
$.0068$ (4B GSM8K), $.0068$ (8B GSM8K), $.0162$ (4B ARC-Challenge),
and $.0213$ (8B MedQA). The pre-registered analysis remains Holm
correction within backbone-defined families; this pooling is a
descriptive companion.

\paragraph{Statistical unit of the equivalence tests.}
The equivalence analyses take the seed, not the example, as the unit
of inference: each arm-seed's 500 examples aggregate to one accuracy,
and TOST runs on the five seed-level true-minus-deranged contrasts.
The choice follows the design. Within an arm-seed, all 500 examples
are answered by a single stochastic receiver realization under one
fixed batch geometry, so examples are not independent replicates of
the receiver-sampling process; the seed is the replication unit. The
per-seed ledger (Table~\ref{tab:seed-td}) shows the realization-level
variance component is real: the true-minus-deranged contrast flips
sign across seeds in seven of the nine natural cells, and the
Nemo-12B GSM8K cell contains a single-seed excursion of $-3.8$ points
against $\approx{-}0.5$ for its other seeds. An example-level paired
analysis pooling 2{,}500 examples per cell would set this component
to zero and report anti-conservatively narrow intervals; a
cluster-robust correction does not escape the problem, since with
seeds as the natural clusters it recovers the same five effective
replicates.

The same logic assigns each cell family its unit. Seed-level
inference is reserved for the equivalence family, where the claim
is a bound. The three-seed cells (calibrated, C2C, KVComm, and the
prose stress test) claim detection at most, and read per-seed
replication --- intersection--union tests where a bound-like
statement is made, descriptive contrasts otherwise; their
batch-level intervals are within-realization precision, not the
inferential unit. The bolt-on cell runs one seed because it is a
specificity control read on an order-of-magnitude contrast
($-36$ to $-46$ against $+1$), not an estimation cell.

The direction of the trade-off matters: the seed-level unit reduces
power, so it makes equivalence harder --- not easier --- to
establish; it can cost equivalence verdicts, as in the
Mistral-Nemo-12B cells, but does not systematically inflate those
reported. The unit was fixed in the analysis plan recorded before
outcome inspection, and switching to a higher-powered unit after
observing which cell missed would be outcome selection; that cell
is accordingly reported exactly as the recorded rule scores it.

\paragraph{What varies across seeds.}
Within a natural cell the 500 examples are a deterministically
selected, frozen subset, identical across seeds and arms; the
derangement is likewise frozen --- a within-batch cyclic shift of
complete cache rows, the same assignment in every seed and arm ---
and the sender-side pipeline (prompt prefill and latent-state
recycling) involves no token sampling, so the relayed caches and the
donor assignment are common to all seeds. A seed varies exactly one
thing: the receiver's sampled decoding. The term
``receiver-sampling seeds'' is literal. The selection and batching
are equally mechanical: the MedQA cell takes the first
500 items under a byte-order sort of source ids, the GSM8K and
ARC-Challenge cells take the first 500 items in dataset order, and
batches are consecutive chunks of twenty in selection order --- no
shuffling and no length bucketing anywhere, so the batch structure
encodes nothing about item content beyond the source ordering
itself.

\paragraph{Coverage of the seed-level interval.}
The equivalence estimand is therefore conditional on the frozen
example set and the frozen donor assignment, and the seed-level
interval covers the one random source that estimand is exposed to:
receiver-decoding noise. Generalization over benchmark items is
carried by replication across cells --- the natural null repeats
across two to three tasks and four backbones --- not by within-cell
item resampling. The fixed derangement, in turn, is not a lone draw
wobbling the estimate: each cell's contrast already averages over
500 distinct donor--recipient pairings, and the marginal-preserving
property that licenses the deranged arm holds for every derangement,
not for a lucky one.

\paragraph{Hierarchical bootstrap companion.}
As a descriptive robustness check that uses the full paired
per-example data, we report a hierarchical bootstrap of each natural
cell's true-minus-deranged contrast: each replicate resamples the
cell's seeds with replacement and, within every drawn seed,
resamples the 500 paired example outcomes with replacement
($10^5$ replicates; percentile 95\% intervals;
Table~\ref{tab:boot}). The bootstrap reproduces
the licensed structure qualitatively: the one detected advantage
stays strictly positive (Qwen3-4B GSM8K, $[+0.32, +2.72]$), the
remaining Qwen3 intervals straddle zero and stay inside the
operationalized margin, and only the two Mistral-Nemo-12B intervals
extend past it ($-3.16$ at the lower end) --- the same two cells the
pre-registered readout scores as equivalence not established.

\begin{table}[t]
\centering
\small
\setlength{\tabcolsep}{4.5pt}
\begin{tabular}{llrr}
\toprule
Cell & Task & Mean T$-$D & Bootstrap 95\% CI \\
\midrule
Qwen3-4B  & GSM8K & $+1.52$ & $[+0.32, +2.72]$ \\
Qwen3-4B  & ARC-C & $+0.20$ & $[-1.68, +2.12]$ \\
Qwen3-8B  & GSM8K & $0.00$  & $[-1.08, +1.00]$ \\
Qwen3-8B  & ARC-C & $-0.36$ & $[-1.24, +0.56]$ \\
Qwen3-8B  & MedQA & $+0.40$ & $[-1.64, +2.36]$ \\
Qwen3-14B & GSM8K & $-0.44$ & $[-1.24, +0.32]$ \\
Qwen3-14B & ARC-C & $-0.04$ & $[-0.76, +0.72]$ \\
Nemo-12B  & GSM8K & $-1.12$ & $[-3.16, +0.80]$ \\
Nemo-12B  & ARC-C & $-0.88$ & $[-3.16, +1.48]$ \\
\bottomrule
\end{tabular}
\caption{Hierarchical (seed $\times$ paired-example) bootstrap of
the true$-$deranged contrast, percentage points; $10^5$ replicates,
percentile intervals. Descriptive companion to the seed-level TOST;
no equivalence verdict is derived from these intervals.}
\label{tab:boot}
\end{table}

\paragraph{Replication on the official GSM8K test split.}
The no-pairing null replicates on the official GSM8K test split, on
items the audited system itself reports on. The companion cell's
design, recorded before any production outcome existed, takes the
first 500 official-test items in dataset order under the same
selection rule as the training-split cell, Qwen3-8B, true and
deranged arms only, the original cell's five receiver seeds
(42--45, 47), the fixed $25{\times}20$ geometry and within-batch
derangement, and the original cell's L40S venue. The readout is
descriptive: per-seed true and deranged accuracies and the
seed-mean true-minus-deranged contrast, reported alongside the
training-split cell's values, opening no new TOST family.

The cell completed under that rule. Per seed (true/deranged,
$n{=}500$ each): $94.2/94.0$, $94.8/94.0$, $93.0/95.0$,
$94.0/94.8$, $93.2/94.6$ percent --- seed contrasts $+0.2$,
$+0.8$, $-2.0$, $-0.8$, $-1.4$ points, sign-mixed, with seed-mean
true-minus-deranged $-0.64$ points. The official-test contrast
sits where the training-split contrast sits (seed-mean $0.00$,
signs mixed), weakening the item-memorization explanation of the
training-split result.

\section{Per-Seed Natural-Regime Ledger}
\label{app:seed-ledger}

Table~\ref{tab:seed-td} reports the raw true-arm and deranged-arm
accuracies behind the natural-regime cell means and confidence
intervals of Table~\ref{tab:natural}; per-seed
true-minus-deranged differences are these column-wise contrasts, and
the per-seed circles of Figure~\ref{fig:dissociation} are drawn from
them. All entries are exact
percentages from $n{=}500$ per arm-seed; the Qwen3-8B GSM8K and
ARC-Challenge cells use seeds 42--45 and 47 (the original seed 46
hit the deterministic out-of-memory failure on their L40S venue and
was replaced wholesale by seed 47 per the replacement rule recorded
in advance --- see Appendix~\ref{app:reproducibility}); the MedQA
cell, which ran on A100-SXM4-80GB, and all other cells use seeds
42--46.

Table~\ref{tab:seed-zr} extends the ledger to the two liveness
arms: per-seed zero-arm and random-arm accuracies for the same
cells, the values behind the natural-regime true-minus-zero and
true-minus-random contrasts quoted in the body. Both corruptions
land generation at the same $\approx\!1\%$ level, so the two
contrasts read one collapse against two corruptions --- which is
what establishes causal coupling at the intervention point.

\begin{table}[t]
\centering
\small
\setlength{\tabcolsep}{6pt}
\begin{tabular}{llccccc}
\toprule
Cell & Task & \multicolumn{5}{c}{True / Deranged accuracy (\%) by seed} \\
\midrule
Qwen3-4B  & GSM8K & 90.8/89.4 & 91.2/89.6 & 91.0/89.2 & 90.8/88.8 & 90.0/89.2 \\
Qwen3-4B  & ARC-C & 87.4/88.6 & 88.2/89.4 & 89.2/87.4 & 88.2/86.6 & 88.4/88.4 \\
Qwen3-8B  & GSM8K & 94.0/93.6 & 94.0/93.0 & 92.6/94.2 & 94.0/93.4 & 93.8/94.2 \\
Qwen3-8B  & ARC-C & 96.8/97.6 & 96.4/97.4 & 96.8/96.0 & 96.2/97.0 & 96.8/96.8 \\
Qwen3-8B  & MedQA & 75.8/74.6 & 73.8/75.8 & 76.2/73.8 & 74.6/75.6 & 75.6/74.2 \\
Qwen3-14B & GSM8K & 94.0/94.4 & 93.4/93.0 & 94.0/94.2 & 92.8/94.2 & 93.6/94.2 \\
Qwen3-14B & ARC-C & 96.4/96.8 & 96.0/96.6 & 96.2/96.6 & 96.6/96.4 & 96.8/95.8 \\
Nemo-12B  & GSM8K & 79.8/80.2 & 81.4/81.8 & 78.2/78.8 & 77.8/81.6 & 78.8/79.2 \\
Nemo-12B  & ARC-C & 13.6/15.4 & 13.6/15.8 & 13.6/12.8 & 14.0/12.4 & 12.4/15.2 \\
\bottomrule
\end{tabular}
\caption{Per-seed true and deranged arm accuracies for all natural
cells. Seeds are in ascending order per cell (42--46, except
Qwen3-8B GSM8K/ARC-C: 42--45, 47). Nemo-12B ARC-Challenge absolute
levels are
interpretability-limited (parse failures scored wrong under the frozen
convention, pushing all arms below four-choice chance); within-cell
contrasts remain valid because the convention is uniform across arms.}
\label{tab:seed-td}
\end{table}

\begin{table}[t]
\centering
\small
\setlength{\tabcolsep}{6pt}
\begin{tabular}{llccccc}
\toprule
Cell & Task & \multicolumn{5}{c}{Zero / Random accuracy (\%) by seed} \\
\midrule
Qwen3-4B  & GSM8K & 92.0/0.4 & 91.8/1.0 & 91.6/1.6 & 92.4/1.8 & 92.2/1.6 \\
Qwen3-4B  & ARC-C & 95.6/0.0 & 95.8/0.0 & 96.0/0.0 & 95.4/0.0 & 96.0/0.0 \\
Qwen3-8B  & GSM8K & 92.8/0.0 & 92.8/0.0 & 93.2/0.0 & 92.8/0.0 & 93.2/0.4 \\
Qwen3-8B  & ARC-C & 96.4/0.2 & 96.0/0.0 & 96.4/0.2 & 96.2/0.2 & 96.2/0.0 \\
Qwen3-8B  & MedQA & 61.6/0.0 & 58.6/0.0 & 58.6/0.0 & 60.4/0.0 & 63.4/0.0 \\
Qwen3-14B & GSM8K & 93.2/0.2 & 93.6/0.0 & 93.0/0.2 & 93.0/0.0 & 93.0/0.0 \\
Qwen3-14B & ARC-C & 96.8/0.0 & 96.2/0.0 & 96.8/0.0 & 96.4/0.0 & 96.2/0.0 \\
Nemo-12B  & GSM8K & 1.6/1.0  & 0.6/1.4  & 0.6/1.8  & 1.6/0.8  & 1.0/1.6 \\
Nemo-12B  & ARC-C & 0.4/0.0  & 0.0/0.0  & 0.6/0.0  & 0.2/0.0  & 0.4/0.0 \\
\bottomrule
\end{tabular}
\caption{Per-seed zero-arm and random-arm accuracies (zero/random,
\%) for all natural cells, recomputed read-only from the raw
per-example ledgers; seeds ascending per cell as in
Table~\ref{tab:seed-td}. These arms carry the quoted
true-minus-zero and true-minus-random contrasts; the random arm's
collapse in every cell shows the intervention reaches the
receiver's computation.}
\label{tab:seed-zr}
\end{table}

\section{Mechanistic and Cross-Surface Evidence}
\label{app:mechanistic}

\subsection{At the Informed-Row Ceiling, Cache Pairing Produces No
Measurable Accuracy Contrast}
\label{app:factorial}

A within-task factorial cell tests moderation directly on a single
surface: it crosses receiver need with relay-cache source while
holding task, prompt template, cache geometry, and model
priors fixed. That removes the
task-distribution difference the cross-regime dissociation carries:
the audit's two regimes are defined by receiver necessity but
realized on different task surfaces, so across regimes necessity is
observational and remains a moderator hypothesis.

\paragraph{Design and result.}
The completed factorial crosses terminal explicit access to the true
registry (absent versus present, the latter via the same judger-only
text injection as the direct-text arm) with the relay-cache source
(aligned true-private versus answer-irrelevant intact-no-private) on
the primary Qwen3-8B interface: the same 500 calibration examples,
batch assignments, and receiver seeds 42--44 as the production cell,
with the new informed-aligned arm's pre-intervention caches matching
the production true-private cell batch by batch. The relay-cache effect within each row,
and the interaction between rows, follow the recorded rule
(batch-level paired contrasts, 25 batches, one-sample $t$). In the
uninformed row the relay-cache effect is large in every seed
($+0.748$, $+0.756$, $+0.766$; 95\% CIs $[0.702, 0.794]$,
$[0.708, 0.804]$, $[0.728, 0.804]$). In the informed row both cells
sit at the $500/500$ ceiling in all three seeds, so the informed-row
relay effect is exactly zero and the interaction numerically equals
the uninformed-row effect. Cache pairing moves accuracy only where
the receiver lacks the content --- but because the informed row is
at ceiling, the interaction estimate degenerates to the
uninformed-row contrast; per the rule recorded at the cell's
design, the factorial supports moderation without establishing it,
now within one task surface rather than across two.

\paragraph{Deranged-cache completion.}
A completion arm crosses the deranged cache with the visible
registry (seeds 42--44, same examples and batches), making the
informed row's contrast the literal true-minus-deranged. The arm
reads $500/500$ in every seed --- identical to its aligned-cache
counterpart --- so the informed-row true-minus-deranged contrast is
exactly zero in all seeds: a content-mismatched full cache produces
no detected interference when the receiver holds the content as
text. The literal-contrast interaction accordingly degenerates, as
its aligned-cache counterpart did, to the uninformed row, where
true-minus-deranged reads $+0.894$, $+0.880$, and $+0.894$ (95\%
CIs $[0.865, 0.923]$, $[0.853, 0.907]$, $[0.861, 0.927]$; per seed,
no pooling). Because the informed row is at ceiling here too, this
contrast likewise supports but does not establish moderation.

\subsection{Donor-Consistent Answers Under Mismatched Relays}
\label{app:donor}

The body reports that under deranged relays in the calibrated
cells, receiver outputs shift toward the donor example's registry.
The analysis definition: an output counts
as a donor match when it parses to the correct choice of the
\emph{donor} example's registry --- the registry whose cache the
derangement delivered --- under the same strict parser as the
accuracy readout; parse failures count as non-matches. The donor
assignment is read from the ledgers' per-example donor identity and
validated against the recorded within-batch derangement in every
batch. Per seed, on the primary interface: under
constrained-deranged relays $321/500$ ($64.2\%$), $324/500$
($64.8\%$), and $323/500$ ($64.6\%$) of outputs match the donor's
label for seeds 42--44 respectively, versus $117/500$ ($23.4\%$),
$106/500$ ($21.2\%$), and $104/500$ ($20.8\%$) under the matched
answer-irrelevant control. The contrast is licensed for the primary
interface only; the latent-only interface reads
$\approx\!3.5\%$ on both arms and supports no such analysis.

\subsection{Surface-Form Stress Test on the Prose Carrier}
\label{app:prose-native}

The body's surface-form stress test reports the ceiling transfer;
this appendix carries the per-arm parse and cap-hit structure behind
it (Table~\ref{tab:prose-native}). The paired-advantage bounds also
pass under the worst-case parse convention, which rescores every
treatment-arm parse failure as wrong and every control-arm parse
failure as correct --- the most adverse resolution of unparsed
outputs for the claimed advantage.

The parse column separates the control arms' failure modes. The
deranged relay leaves the answer format intact (parse
$97.8$--$98.8\%$) while the answers it produces are wrong ---
correctness among its parsed outputs runs $5$--$6\%$, far below
the $25\%$ four-choice chance its parse rate would support,
consistent with the donor-following behavior measured directly on
the registry surface. The intact and zero arms instead lose the
answer format (parse $\approx\!18$--$26\%$) and sit at chance
among the outputs that do parse. The asymmetry is a property of
the delivered native relay, not of the carrier: in the C2C cell
on this same prose surface (Table~\ref{tab:c2c-prose}), all three
projected-cache arms parse at $\approx\!50\%$ with no arm
asymmetry.

Three reference levels were fixed before the audit ran: $25\%$ is
nominal four-choice chance; $28.6\%$ is the strongest audited
uninformed heuristic --- the frozen construction probes (a
feature-based edited-slot classifier and a static slot-to-label
majority rule) each reach $143/500$; and $31\%$ is the
pre-registered interpretation cutoff, frozen as a constant in the
floor-audit code before the audit ran: per-seed intact-no-private
and zero accuracies must stay at or below it for the ceiling
reading to be licensed. It is a licensing threshold, not a
mathematical upper bound on every uninformed strategy. Every
control arm sits far below all three levels
(Table~\ref{tab:prose-native}).

\begin{table}[t]
\centering
\small
\setlength{\tabcolsep}{5pt}
\begin{tabular}{lccccccccc}
\toprule
Arm & \multicolumn{3}{c}{Correct / 500} &
\multicolumn{3}{c}{Parse rate (\%)} &
\multicolumn{3}{c}{Cap-hit rate (\%)} \\
\cmidrule(lr){2-4}\cmidrule(lr){5-7}\cmidrule(lr){8-10}
 & 42 & 43 & 44 & 42 & 43 & 44 & 42 & 43 & 44 \\
\midrule
True             & 500 & 500 & 500 & 100  & 100  & 100  & 0    & 0    & 0    \\
Direct-text      & 500 & 500 & 500 & 100  & 100  & 100  & 0    & 0    & 0    \\
Intact           & 22  & 31  & 20  & 17.8 & 22.4 & 20.0 & 26.4 & 26.4 & 24.8 \\
Deranged         & 31  & 31  & 26  & 97.8 & 98.8 & 98.4 & 0.8  & 1.2  & 1.0  \\
Zero             & 30  & 23  & 33  & 24.2 & 25.8 & 23.0 & 75.8 & 74.2 & 77.0 \\
\bottomrule
\end{tabular}
\caption{Native-relay prose-carrier cell (Qwen3-8B), per seed:
correct answers out of 500, parse rate (share of outputs parsing
to a choice under the strict frozen parser; parse failures score
wrong), and generation-cap hit rate. All control arms sit below
the pre-registered $31\%$ interpretation cutoff in every seed.}
\label{tab:prose-native}
\end{table}

\subsection{Replication on a Prose Carrier (C2C)}
\label{app:c2c-prose}

The C2C detection-level null replicates on the fixed-frame
constructed-prose carrier, weakening surface-form mismatch as an
explanation of the registry result. The pre-registered extension
cell reruns the C2C battery on the prose counterfactual QA surface
(the carrier of Appendix~\ref{app:construction}) because the
registry-surface null admits two readings: C2C's fusers are trained
on natural-language corpora \citep{fu2025_cache}, so the registry
format may fall outside the projector's training distribution.

The cell's design and readout rule, recorded before any production
outcome existed: three receiver seeds (42--44), the registry-surface
C2C cell's five arms unchanged (true-private, constrained-deranged
private, intact-no-private, zero, and direct-text reference; the
moment-matched random arm belongs to the native-relay battery and
was never part of the delivered C2C cell), $n{=}500$ prose
counterfactual QA examples in the fixed $25{\times}20$ batch
geometry, Qwen2.5-7B sender into Qwen3-8B receiver on
A100-SXM4-80GB --- the registry-surface C2C cell's venue and
statistics unchanged. The readout is descriptive and
detection-level, exactly as in the registry C2C cell: per-seed arm
accuracies and arm contrasts, no TOST family and no equivalence or
bounded wording. A persistent null (true and deranged
indistinguishable at detection level in every seed) weakens the
out-of-distribution-form explanation of the registry-surface null;
a detected true-minus-deranged separation is reported as a boundary
finding about the projector's input distribution. A tokenizer
geometry gate preceded production: all 500 prose carriers render to
identical token identities, active lengths, and
relation/query/option positions across the sender and receiver
tokenizers, with target-only edit containment and intact
within-batch donor geometry.

\paragraph{Result.}
The pairing contrast (true-private minus constrained-deranged) is
sign-mixed across seeds ($-0.2$, $-0.2$, $+2.6$ points;
Holm-adjusted $p=.91$): no detected example-specific transfer
(Table~\ref{tab:c2c-prose}). The recorded analysis tests the
true-versus-intact comparison as the cell's primary readout (a
three-seed intersection--union test) and places the five secondary
accuracy contrasts --- T$-$D, T$-$Z, D$-$I, I$-$Z, and
direct-text$-$I --- in one Holm family, all five reported here:
T$-$D $p=.91$, T$-$Z $p=.001$, I$-$Z $p<.001$, D$-$I $p=.84$ (no
directional reading licensed), direct-text$-$I $p<.001$. The cell is
internally live on the same
evidence pattern as the registry cell: the zero arm falls below
every projected-cache arm (true minus zero $+5.4$/$+6.8$/$+7.6$
points; intact-no-private minus zero
$+7.0$/$+8.6$/$+8.6$ --- the generic term is
detected while the pairing term is not, the registry cell's
decomposition structure), and the direct-text arm confirms task and
parser. On the primary true-versus-intact comparison the point
estimate favors intact in all three seeds ($-1.6$/$-1.8$/$-1.0$
points; intersection--union $p=.48$): at three seeds the cell
distinguishes neither direction, so the licensed status is no
detected separation between the true and matched-control arms ---
not equivalence, which this detection-level cell does not claim. Projected-cache
arms carry heavy truncation and parse loss on this surface
(Table~\ref{tab:c2c-prose}), so absolute levels are
interpretability-limited while within-cell contrasts remain valid
--- the same convention as the Nemo-12B ARC-Challenge cell.

\begin{table}[t]
\centering
\small
\setlength{\tabcolsep}{5pt}
\begin{tabular}{lccccccccc}
\toprule
Arm & \multicolumn{3}{c}{Correct / 500} &
\multicolumn{3}{c}{Parse rate (\%)} &
\multicolumn{3}{c}{Cap-hit rate (\%)} \\
\cmidrule(lr){2-4}\cmidrule(lr){5-7}\cmidrule(lr){8-10}
 & 42 & 43 & 44 & 42 & 43 & 44 & 42 & 43 & 44 \\
\midrule
True         & 60  & 56  & 64  & 49.8 & 51.0 & 50.2 & 39.4 & 35.8 & 35.2 \\
Deranged     & 61  & 57  & 51  & 55.8 & 49.4 & 51.0 & 34.4 & 37.0 & 36.0 \\
Intact       & 68  & 65  & 69  & 53.2 & 49.8 & 53.0 & 34.2 & 37.2 & 36.0 \\
Zero         & 33  & 22  & 26  & 18.6 & 17.4 & 19.0 & 66.4 & 71.0 & 65.8 \\
Direct-text  & 495 & 490 & 490 & 99.0 & 98.4 & 98.2 & 0.4  & 0.6  & 1.6  \\
\bottomrule
\end{tabular}
\caption{C2C prose-surface cell, per seed: correct answers out of
500, parse rate, and generation-cap hit rate per arm. The three
projected-cache arms cluster together on all three readouts; under
the zero arm, parse rate falls and cap-hit rate rises sharply ---
the total cache contrast separates a well-formed projected cache
from a degenerate one on this surface too.}
\label{tab:c2c-prose}
\end{table}

\subsection{Scope of the Cross-System Comparison}
\label{app:cross-system-scope}

The three-level reading (ceiling, partial, none detected) is a
statement about three delivered configurations under one
construction and one receiver family, not a capacity ranking of
three communication architectures. For placement: KVComm's smallest
per-seed pairing contrast under the same construction and receiver
family ($+7.2$) exceeds C2C's largest ($+2.0$) by more than a factor
of three, and KVComm's contrasts are positive in every seed while
C2C's change sign.

For C2C, integration liveness is internal to the cell: the zero arm
collapses to $\approx$6\% while every projected-cache arm clusters
near 22\%, so the delivered projector demonstrably changes receiver
behavior relative to a geometry-matched empty cache --- a broken
integration could not produce that separation. The direct-text arm
validates task and parser independently. C2C's main
pairwise-performance tables fix the receiver at Qwen3-0.6B, and the
audited Qwen2.5-7B-Instruct$\to$Qwen3-8B configuration appears in
C2C's published evaluation only as a single untabulated point in its
model-size scaling figure (MMLU-Redux, a small gain, consistent with
that paper's own observation that gains shrink for larger
receivers), with no evaluation recipe shipped for the pair. The
audited cell therefore makes no contact with any tabulated C2C gain;
the pair was selected because its receiver matches the audit's
primary Qwen3-8B backbone, keeping the cross-system comparison
within one receiver. A reproduction of a C2C headline benchmark
result would validate the released weights in a different,
small-receiver regime; it is complementary to, not a precondition
for, auditing the delivered artifact's behavior under an
information-asymmetry probe.

Inference levels are deliberately matched to claims. The natural
cells carry TOST machinery because equivalence is claimed there; the
C2C and KVComm cells claim detection-level readings only ---
``sign-mixed'', ``no detected example-specific transfer'',
``detected, far below ceiling'' --- and three seeds of descriptive
contrasts license exactly those words. No bounded or equivalence
wording is used for either ported system.

\section{Decomposing Relay Effects}
\label{app:decomposition}

\subsection{Receiver-Only Baselines}
\label{app:norelay}

A receiver-only condition --- the frozen example sets, prompts, and
decoding seeds with no cache relay at all --- separates two readings
of the zero arm's cost that the four delivered arms cannot: whether
zeroing removes genuinely useful generic state or damages the
receiver through the retained interface. No-relay accuracy near the
zero arm reads the zero arm as removing real generic value;
no-relay accuracy near the true arm, with zero below both, reads
the zero arm's cost as interface damage.

The cell's design and comparison rule, recorded before any
production outcome existed: receiver-only generation --- no cache
prefix, positions from zero, plain attention mask --- on the frozen
example sets, prompts, and decoding seeds of the corresponding
delivered cells, at the same venue class per cell
(Table~\ref{tab:venues}). The readout is descriptive estimation
against the existing arm means --- per-seed no-relay accuracy placed
against the cell's true, deranged, and zero arm values --- under the
two-reading rule above; it opens no TOST family, since no pre-set
margin exists for a receiver-only contrast.

\begin{figure}[t]
\centering
\includegraphics[scale=1]{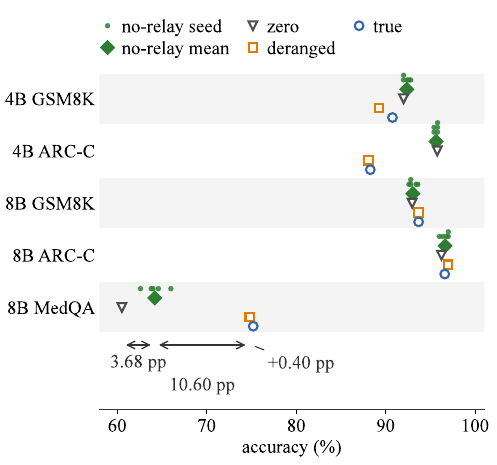}
\caption{\textbf{Receiver-only placement.} Receiver-only (no-relay)
accuracy tracks the zero arm on the four Qwen3 GSM8K/ARC-Challenge
cells, and at 4B the delivered arms sit below the receiver-only
level; the MedQA row splits its total cache contrast into interface
damage (zero to no-relay), generic value (no-relay to deranged),
and the pairing contrast (deranged to true). Per-seed no-relay
accuracies (dots; exact duplicates stacked; filled diamond at the
mean) are placed against the delivered arm means, one lane per
condition within each row so coincident values stay individually
visible. The readout is descriptive and opens no test family.}
\label{fig:norelay}
\end{figure}

\paragraph{Result.}
Receiver-only accuracy tracks the zero arm on the four Qwen3
GSM8K/ARC-Challenge cells; MedQA alone shows a real zero-arm cost
(Table~\ref{tab:norelay}, Figure~\ref{fig:norelay}). All five
prespecified cells completed at their delivered cells' venue class,
seed set (42--46; Qwen3-8B GSM8K/ARC: 42--45 and 47), example sets,
prompts, and parser convention, with $n{=}500$ per seed and zero
relay events recorded in every job.

On the four Qwen3 cells the seed-mean no-relay-minus-zero
differences are $+0.36$, $-0.12$, $+0.08$, and $+0.40$ points, so
the retained zeroed interface reads as undamaging on these
surfaces. At 8B on GSM8K and ARC-Challenge the true arm sits at
the same level and there is no zero-arm cost to decompose; the
MedQA cost is decomposed below. At 4B the
delivered arms sit below the receiver-only level in every seed pair
--- seed-mean $-1.60$ points against the true arm on GSM8K and
$-7.36$ on ARC-Challenge --- which reads the relay's generic effect
on these two surfaces as a cost the receiver avoids by solving
alone, while the pairing contrast between the delivered arms stays
at the bounded near-zero values of the main analysis.

\paragraph{Absolute levels against the source report.}
The audited true arm's absolute accuracies sit close to the
sequential accuracies the audited system itself tabulates
\citep[Table~1]{zou2025_latent}. Per cell
(audited$-$tabulated): 8B GSM8K $-0.12$, 8B MedQA $-0.10$, 4B GSM8K
$+2.56$, 4B ARC-Challenge $-4.02$, 8B ARC-Challenge $+2.20$, 14B
GSM8K $-1.64$, 14B ARC-Challenge $+0.80$ --- mixed-sign divergences
within about four points across seven cells, two matching to
within $0.12$, the pattern a reproduction run on a different
500-item subset produces. The comparison places absolute levels;
it cannot by itself validate implementation fidelity, since the
audited item sets differ from the source's. The receiver-only
condition is not the system's ``Single'' baseline: it removes the
relay from the receiver role while keeping the role's prompt
surface, so its placement above the delivered arms at 4B is a
reading of the relay's generic cost on those surfaces, not a
reproduction failure. On training-split items, memorization raises
unaided-receiver accuracy specifically --- the same direction
already disclosed for the equivalence bound.

The receiver-only condition splits MedQA's zero-arm cost: no-relay
($64.20\%$) sits $3.68$ points above the zero arm (positive in all
five seeds), $10.60$ points below the deranged arm, and $11.00$
points below the true arm, reading the $14.68$-point total cache
contrast as $3.68$ points of interface damage from the retained
zeroed geometry, $10.60$ points of generic value carried by a
delivered cache regardless of pairing, and the $+0.40$-point pairing
contrast between the delivered arms
(Table~\ref{tab:natural}). A parse-channel
count locates part of that generic value: 554 of 2500 receiver-only
generations fail the frozen answer parser, against 34 under the
delivered true cache and 660 under the zero arm, so part of the
generic value on this surface is compliance with the answer format
rather than task content.

\paragraph{Zero-arm semantics.}
An all-zero KV block is itself outside the receiver's training
distribution; for exactly this reason the true-minus-zero quantity
carries the name \emph{total cache contrast} --- a contrast between
two delivered conditions. Any
structural penalty the zeroed cache imposes beyond removing content
inflates the generic terms ($\tau_{IZ}$, $\tau_{DZ}$) and leaves the
content terms ($\tau_{TI}$, $\tau_{TD}$) untouched; under that
reading the generic terms are upper bounds on generic-computation
value, the calibrated dominance of the content term is conservative,
and the natural-regime pairing bound does not involve the zero arm
at all. The data above show directly that the total cache contrast
is not a stable generic-value measurement: on the natural surfaces
the true-minus-zero contrast runs from $-1.2$ and $-7.5$ points on
the Qwen3-4B cells through
no visible cost at 8B on GSM8K and ARC-Challenge to $+14.7$ points
on MedQA and $+78.1$ points on Nemo-12B GSM8K ---
changing sign and spanning near zero to $+78$ points across receivers
running the same delivered relay while the pairing term stays near
zero in every one of those cells, bounded in the Qwen3 cells and
undetected on Nemo-12B. That
receiver-idiosyncrasy is consistent with an interface-level effect
and is why the decomposition's generic terms are reported as
contrasts, not benefits. The cleaner generic control --- a
same-example, answer-neutral relay --- exists only where
sender-private content exists to withhold: that arm is the
calibrated regime's intact-no-private, and it cannot be constructed
on the natural surfaces, where the sender processes nothing the
receiver lacks.

\begin{table}[t]
\centering
\small
\setlength{\tabcolsep}{6pt}
\begin{tabular}{llcrrrr}
\toprule
Cell & Task & No-relay by seed (\%) & Mean & True & Der. & Zero \\
\midrule
Qwen3-4B & GSM8K & 92.0 / 92.6 / 92.0 / 92.8 / 92.4 & 92.36 & 90.76 & 89.24 & 92.00 \\
Qwen3-4B & ARC-C & 95.4 / 95.4 / 95.8 / 95.8 / 95.8 & 95.64 & 88.28 & 88.08 & 95.76 \\
Qwen3-8B & GSM8K & 92.8 / 92.8 / 92.6 / 93.4 / 93.6 & 93.04 & 93.68 & 93.68 & 92.96 \\
Qwen3-8B & ARC-C & 96.0 / 96.4 / 96.8 / 97.0 / 97.0 & 96.64 & 96.60 & 96.96 & 96.24 \\
Qwen3-8B & MedQA & 64.0 / 63.8 / 64.6 / 62.6 / 66.0 & 64.20 & 75.20 & 74.80 & 60.52 \\
\bottomrule
\end{tabular}
\caption{Receiver-only (no-relay) per-seed accuracies with the
corresponding delivered-cell arm means (true, deranged, zero) for
placement. Each no-relay cell reuses its delivered cell's frozen
example set, prompts, decoding seeds, venue class, and parser
convention; seeds ascending per cell (42--46, except Qwen3-8B
GSM8K/ARC: 42--45, 47).}
\label{tab:norelay}
\end{table}

\subsection{KVComm: Detected Value Is Pairing-Specific}
\label{app:kvcomm-zero}

In the KVComm cell the total cache contrast is no larger than the
content contrasts (true$-$zero $+7.8/{+}6.8/{+}7.4$ points per seed
against true$-$deranged $+10.8/{+}7.2/{+}9.0$), so the deranged,
intact-no-private, and zero arms sit level as a group. Read through
the decomposition, the generic-computation term does not separate
from zero in this cell while the pairing term does: the relay's
detected value sits in the pairing term. That differs from the other audited
systems --- LatentMAS's calibrated cells carry a substantial generic
term (intact well above zero), and C2C's total cache contrast exceeds
its content contrasts. That variation across delivered systems is
itself a reading a presence-only instrument could not produce.

Layer selection is calibrated once and held fixed
across all five arms, so within-cell contrasts are conditional on the
selector but not confounded by it. The calibration is task-matched
--- a setting favorable to KVComm --- and the layer-selected relay
still recovers roughly a tenth of the full-prepend ceiling.

\section{Identification Detail and Boundaries}
\label{app:identification}

\subsection{Marginal Preservation and Donor Collisions}
\label{app:donor-similarity}

The within-batch derangement is marginal-preserving by construction,
so on homogeneous task surfaces donor-similarity effects --- a donor
cache from a similar problem carrying reasoning content that
transfers to the queried problem --- reduce to the generic term the
decomposition already measures. Four observations place that
reduction on the data.

First, transferable content is outside the pairing estimand by
construction. Pairing value is what correct cache-to-query assignment
adds over a mismatched assignment; whatever a mismatched donor
supplies --- task format, generic reasoning scaffolding --- is the
generic term $\tau_{DZ}$ of the decomposition, which the audit
measures and reports separately rather than denying. Where measured
it is substantial: on MedQA zeroing the relay costs the
receiver 14.7 points while the pairing term reads $+0.4$, and the
Nemo-12B total cache contrast reaches $+78.1$ points on GSM8K.
Reading that transfer as dilution of the pairing effect relabels
the generic term as pairing; the
attribution under audit is the pairing term.

Second, a maximally-dissimilar donor arm would break identification
rather than sharpen it. The derangement is constrained within batch
precisely because it is marginal-preserving: both arms deliver the
same multiset of caches and only the assignment changes, so a
contrast cannot be produced by malformedness or distribution shift.
A dissimilar donor (a cross-task or cross-distribution cache)
reintroduces distribution shift, and its contrast would read pairing
plus shift with no way to separate the two. The existing arms already
bracket the endpoints: the zero arm removes content wholesale while
retaining cache geometry, and the moment-matched random arm shows
what distribution-violating content does --- accuracy collapses.

Third, the convergence of true and deranged arms in the natural
regime is not receiver blindness to the relay: under calibrated
derangement (primary interface), receiver outputs track the
\emph{donor's} registry label ($64.2$--$64.8\%$ versus
$20.8$--$23.4\%$ under the matched control), so the receiver
demonstrably reads mismatched content when content matters.

Finally, the fixed derangement's donor--recipient similarity can be
bounded directly at the answer level. Under the recorded
within-batch assignment, the donor's gold answer coincides with the
recipient's on $7/500$ examples ($1.4\%$) in every GSM8K cell:
almost no deranged cache carries the queried answer, and the
true-minus-deranged contrast stays near zero all the same, so
the natural null is not donor caches smuggling in correct answers.
On the multiple-choice surfaces the collision rate sits near the
label-frequency baseline ($132/500 = 26.4\%$ on ARC-Challenge;
$112/500 = 22.4\%$ on MedQA) --- four-option
surfaces cannot fall below chance-level collision by construction
--- and the same null holds there. The two surface classes bracket
the collision range, and the contrast does not move with it.

\subsection{Full-Cache Estimand and $\tau_{TI}$ Semantics}
\label{app:full-cache}

The audit intervenes on the full relayed cache --- role-prompt,
input-token, and latent-position KV entries with their position and
mask structure, keys and values jointly --- because the full cache
is the delivered mechanism: the audited system relays it whole, its
information-preservation theorem states the attribution at exactly
this level, and the estimand is named accordingly (pairing value of
the relayed cache, never ``latent-thought transfer''). In the
natural regime the derangement swaps the complete post-sender relay
state, so the bounded total is the sum of all components'
example-specific contributions; components could strictly offset
within a bounded total, but that reading requires near-exact
cancellation to recur across tasks, scales, and seeds, and the
audit reports the deployment-relevant total. A position-partition
factorial (deranging latent-generated against input positions) is
the natural mechanism-attribution follow-up; the audit's claims are
stated at the level of the delivered relay.

\paragraph{The causal semantics of $\tau_{TI}$.}
The intact-no-private intervention operates on the sender's
\emph{input}, so the sender's entire internal computation shifts
with it: $\tau_{TI}$ identifies the effect of the full sender-state
difference induced by replacing the private content, which the
matched construction makes plausibly binding-dominated but does not
isolate surgically. The pairing estimands the audit's claims rest
on are free of this caveat by construction: the derangement behind
$\tau_{TD}$ leaves every sender computation untouched and reassigns
only the delivered caches.

\subsection{Intervention Boundary in the Audited Chain}
\label{app:chain-boundary}

The calibrated cells intervene at the final relay boundary of the
four-role chain: the cache delivered to the judger. That cache is
the accumulated post-sender state --- in the primary interface the
planner's prompt and latent positions survive physically through
the critic and refiner stages (the judger receives a 544-row
physical cache whose semantic positions span the planner prompt at
0--179, its latent block at 180--219, and the later roles'
contributions), so the intervention swaps the chain's cumulative
working state, not one hop's increment. The natural cells are
single-hop sender-to-receiver relays, and the intervention boundary
is the one relay they contain.

\subsection{Specificity Control: Bolt-On Latent Prefix}
\label{app:bolton}

The audited systems relay \emph{native} internal state. For contrast,
this pre-registered cell tests a naive constructed channel:
suppressed-token prefix injection, in which a sender's hidden decision
signal is encoded as a fixed-length (16-slot) prefix of non-vocabulary
latent embeddings prepended to a receiver--critic's context. The
cell runs Qwen3-8B drafts against a Qwen3-4B critic, one decoding
seed, batches of five on the shared L40S node, under its own
pre-registered design: four conditions on shared drafts, GSM8K and
ARC-Challenge, 100 examples per task. Net correction counts the
critic's flips relative to the shared drafts --- examples flipped
wrong-to-correct minus examples flipped correct-to-wrong under the
critic's KEEP/REPLACE decision; the lenient parser accepts the
decision line anywhere in the output, and compliance is the share
of outputs emitting a parseable decision line. An output with no
parseable decision line yields no prediction and scores its example
wrong, so format breakage feeds directly into the net-correction
count: the net-correction and compliance readouts are two views of
one injection-damage channel, not independent measurements.

The payload carries no usable example-specific signal, and
injection damages the critic regardless of content. With no
injection the critic's net correction is $+1$ (pooled, lenient
parse); with the same information \emph{visible} as text, $-4$;
with the hidden payload injected, $-36$; with a \emph{shuffled}
payload, $-46$. Hidden and shuffled sit in the same range, far
below the visible-text condition, and format compliance drops from
100\% to 74--81\% under injection. A follow-up
single-arm control replaces the payload with uniform-random token
embeddings (same length, position, and format, sampled seed-fixed from
the vocabulary excluding special tokens): pooled lenient net correction
$-46$ with 74.5\% compliance --- inside the hidden/shuffled range. Per
the pre-registered rule, the damage is attributable to the injection
mechanism itself, not to payload content or its token distribution.

\subsection{Latent-Only Interface Anchor}
\label{app:latent-only}

One empirical anchor for the component question comes from the
delivered system itself. Its own latent-only
(repeated-truncation) interface --- which discards all prompt and
prior-role KV rows after each non-judger role and retains only the
newest latent-generated positions --- was run as a secondary
interface over the same four-arm battery (seeds 42--44, $n{=}500$).
Under it, the true and deranged arms both collapse to
$\approx\!3.5\%$ accuracy against the primary interface's ceiling,
with parse success itself collapsing ($\approx\!19\%$); control runs
attribute the collapse to the upstream interface
semantics rather than to any harness defect. This is consistent
with the registry recovery in the calibrated cells flowing through
retained prompt-KV rows, but the accompanying parse collapse makes it an
interface-level finding, not a clean component attribution, and
the donor-consistency analysis remains licensed for the primary
interface only.

\subsection{Portability Boundary: Llama-3.2-3B}
\label{app:llama}

The Llama family is a portability boundary, not an audit result:
under both delivered realignment configurations the receiver could
not consume the relay, so the audit's precondition of a functioning
channel was not met (Table~\ref{tab:llama}). The direct evidence is
from Llama-3.2-3B-Instruct \citep{meta2024_llama32}; a planned
Llama-3.1-8B cell was closed
without running once both configurations had failed on 3.2-3B. In
both configurations degeneration begins with the latent-feedback
transform, whose norm rescaling collapses hidden states by roughly
$83\times$ (raw norms $90.30$--$90.45$ scaled to a target of
$1.086$); the delivered least-squares realignment matrix leaves
norms essentially unchanged before that same rescaling
($87.4$--$90.5$ before and after the matrix), so it preserves the
collapse and does not restore a consumable relay. Controls isolate
the cause: the stock model, the cache wrapper without injection, and
cache injection with zero latent steps all generate coherently,
while a single latent step per role already breaks generation.

\begin{table}[t]
\centering
\small
\setlength{\tabcolsep}{3.5pt}
\begin{tabular}{p{0.32\textwidth}p{0.56\textwidth}}
\toprule
Configuration & Outcome on Llama-3.2-3B-Instruct \\
\midrule
Delivered default (identity realignment) &
Direct-text-reference arm degenerated to repeated token fragments in
100/100 outputs (smoke cell, 100 examples, 40 latent steps per
role); every output ran to the 4{,}096-token generation cap with a
0\% parse rate; process exited cleanly --- behavioral failure, not a
crash \\
\addlinespace
Learned (least-squares) realignment &
Outputs remain incoherent, 0/3 parseable answers (probe on the same
examples, one latent step per role) \\
\bottomrule
\end{tabular}
\caption{Portability boundary evidence. Both delivered realignment
configurations fail on Llama-3.2-3B-Instruct; the Llama-3.1-8B cell
was closed without running.}
\label{tab:llama}
\end{table}

\subsection{Complementary Estimands: The Audited System's Hybrid
Ablation}
\label{app:hybrid-ablation}

Our natural-regime equivalence bound and LatentMAS's hybrid ablation
(v3, appendix Table~7) answer different questions and can hold
simultaneously.

The ablation's text arm keeps latent reasoning but replaces the KV
relay with decoded text truncated to the last 128 tokens; the reported
drop therefore conflates the communication medium with a hard
compression rule. The audited system's own full-text baseline (TextMAS, text
reasoning \emph{and} text communication) scores \emph{higher} than the
hybrid arm on every reported task (e.g., 92.3 vs.\ 85.5 on GSM8K), so
the ablation cannot be read as text-medium harm per se. The ablation
matches no token or compute budget between arms and does not
specify how latent thoughts are decoded into the transmitted text.

Our audit holds the delivered latent interface fixed and intervenes on
cache \emph{content} within it --- a within-interface estimand that
neither requires nor contradicts any particular latent-versus-text
medium comparison. The reported drop under one specific
truncated-text substitution is compatible with no detected
example-specific value in the relayed content on tasks the receiver
solves unaided.

\section{Construction and Implementation Details}
\label{app:construction}

\subsection{Registry and Prose-Carrier Construction}
\label{app:carrier}

The registry construction is specified with the calibration
instrument in the Audit Design section. The prose carrier used by
the surface-form stress test and the C2C prose-surface cell is a
single prose
paragraph of four templated sentences, one per entity--value
relation: the $k$-th sentence is the $k$-th fixed prefix (``During
the survey,~'', ``Later,~'', ``Meanwhile,~'', ``Finally,~'')
followed by the entity symbol, the fixed relation text `` carried
the signal word~'', the value symbol, and a period. The renderer
enforces paragraph shape and records exact character spans for
every entity and value occurrence, so downstream construction edits
symbols only at recorded spans, never by string search over free
text.

Entity and value symbols are drawn from pools pre-filtered by the
receiver's tokenizer: a candidate symbol is admitted only if it
tokenizes to the same token count inside every frame context it can
occupy. Documents built from admitted symbols share one constant
prompt length, and the per-example token audit verifies that all
arms' prompts have identical token counts, identical special-token
positions, and target-only token edits, so position ids, attention
masks, and cache geometry are constant across arms by construction
rather than by padding. The answer-irrelevant counterpart document
(the intact-no-private analogue on this surface) replaces symbols
at the recorded spans with pool counterparts under the same
geometry constraints, and the same audit gates it. An arm-symmetric
elimination analysis runs on these audited spans; the body's
control-arm caveats are the licensed statement of what that
analysis supports.

\subsection{Task Selection and Feasibility}
\label{app:task-selection}

The natural-regime surfaces are drawn from the audited system's own
nine-benchmark suite, filtered by statistical feasibility: the frozen
design requires $n{=}500$ examples per arm-seed in a fixed
$25{\times}20$ batch geometry, which only four of the nine test sets
support (GSM8K 1{,}319, ARC-Challenge 1{,}172, ARC-Easy 2{,}376,
MedQA 1{,}273; AIME24/25 offer 30 problems each, GPQA-Diamond 198,
HumanEval+ 164, MBPP+ 378). The cells audit three of the four
feasible surfaces, taking the harder of the two ARC variants. The
suite's longer-horizon surfaces are therefore unmeasured for the
pairing term, and the audit's evidence standard applies to them
unchanged: gains are read as transmission only after the same audit.
The deterministic item-selection and batching rules are stated in
Appendix~\ref{app:robustness}, where they support the seed-level
unit.

\subsection{Intervention Mechanics by System}
\label{app:mechanics}

All cells share the arm invariant of the Audit Design section; at
the mechanics level, identical geometry covers layer count,
per-layer sequence length, dtype, position offsets, and
attention-mask structure, with keys and values intervened on
jointly. The delivered mechanism
under intervention differs by family. In the native-relay
(LatentMAS) cells, calibrated and natural, the relayed object is
the full post-sender cache, prepended whole; interventions replace
its content per arm (aligned, deranged reassignment, zeros,
moment-matched noise) while the retained interface keeps cache
lengths, the receiver's position offsets, the attention-mask row,
and zero-valued key slots in the attention normalization. In the
KVComm port, the relayed object is the delivered selector's 11 of
36 layers (layer 0 always transferred in full, the delivered
top-30\% importance rule unchanged); interventions act on the
selected layers' content, the unselected layers remain the
receiver's own computation throughout, and the zero arm zeroes the
selected layers' content while preserving their per-layer lengths
and offsets. In the C2C cell, the relayed object is the released
projector's output cache fused into the receiver; interventions
act on the projected cache (aligned, deranged, zero,
intact-no-private), so every arm delivers a projector-shaped
object and the contrast never compares projected against
unprojected states. In every family the deranged arm reassigns
complete cache objects within batch --- never mixing rows across
examples --- so the delivered multiset is preserved exactly.

\subsection{KVComm Port Mechanics}
\label{app:kvcomm-port}

Beyond the calibration-split replacement disclosed in the body,
the Qwen3-8B port adds a calibration-only Qwen3 attention tracer
(KVComm ships tracers for its original backbones only) and pins the
delivered singleton non-shift-back branch with no forwarded custom
attention mask. The delivered importance statistic, top-30\% rule,
implicit full transfer of layer 0, cache filtering, and
receiver-continuation semantics are unchanged from the released
implementation. The zero arm's retained geometry: per-layer cache
lengths and receiver position offsets are preserved, and zero-valued
key slots remain in the attention normalization, so the ablation
removes cache content without removing the cache interface.

\section{Reproducibility Record}
\label{app:reproducibility}

\subsection{Software Stack, Hardware, and Venues}

Every job of one cell runs on a single GPU model (within-cell
homogeneity); the model varies across cells and is disclosed here
(Table~\ref{tab:venues}). All cells use a pinned software stack
(torch 2.9.0+cu128, transformers 4.57.6, bf16 HF-transformers
generation, no alternative serving stacks). The frozen batch geometry
($25{\times}20$, $n{=}500$ per arm-seed) is never shrunk to fit
memory --- cells that exceed a card's capacity move to a larger card
instead.

\begin{table}[t]
\centering
\small
\setlength{\tabcolsep}{6pt}
\begin{tabular}{llll}
\toprule
Cell & GPU & Seeds & Arms \\
\midrule
Natural 4B / 8B (GSM8K, ARC-C)     & L40S 48GB       & 5 & 4 \\
Natural 14B (GSM8K, ARC-C)         & A100-SXM4-80GB  & 5 & 4 \\
Natural Nemo-12B (GSM8K, ARC-C)    & L40S 48GB       & 5 & 4 \\
Natural 8B (MedQA)                 & A100-SXM4-80GB  & 5 & 4 \\
Calibrated 4B / 8B                 & L40S 48GB       & 3 & 5 \\
Calibrated 14B / Nemo-12B          & A100-SXM4-80GB  & 3 & 5 \\
Calibrated phi-4                   & A100-SXM4-80GB  & 3 & 5 \\
Prose counterfactual QA (8B)       & L40S 48GB       & 3 & 5 \\
C2C (Qwen2.5-7B$\to$Qwen3-8B)      & A100-SXM4-80GB  & 3 & 5 \\
KVComm (Qwen3-8B port)             & L40S 48GB       & 3 & 5 \\
Bolt-on (8B drafts $\to$ 4B critic) & L40S 48GB      & 1 & 4+1 \\
\bottomrule
\end{tabular}
\caption{Per-cell venue and design. All cells use bf16 HF-transformers
generation with a pinned software stack.}
\label{tab:venues}
\end{table}

\paragraph{Compute.}
Experiments ran on two venue classes: a shared 8$\times$L40S 48GB
node, and A100-SXM4-80GB nodes at the Ohio Supercomputer Center
\citep{osc1987} and on a commercial GPU cloud (interchangeable for
within-cell homogeneity as the same GPU model). Per-job runtime
records capture the physical GPU, driver, and library versions.

\subsection{Run Deviations and Resolutions}

The recorded conventions apply uniformly: unconditional accuracy
with parse failures scored wrong, no post-hoc arm acceptance, Holm
correction within pre-registered families, and the TOST margin
recorded before any unblinding. Throughout, ``pre-registered'' and
``recorded in advance'' denote each cell's dated design document ---
arms, validity conditions, margin, hypothesis family, and analysis
rule written down before that cell's outcomes existed --- shipped
verbatim in the release, not a third-party registry entry.
Table~\ref{tab:deviations} records every deviation and
staged-integration event across the battery and its resolution.

\begin{table}[t]
\centering
\small
\setlength{\tabcolsep}{4pt}
\begin{tabular}{p{0.30\textwidth}p{0.28\textwidth}p{0.34\textwidth}}
\toprule
Event & Effect on reported estimand & Resolution \\
\midrule
Natural 4B/8B cells ran initial three-seed cells before the
confirmatory five-seed reruns &
None: the initial runs informed no analysis choice and their numbers
are not reported &
Confirmatory five-seed cells are the reported record \\
\addlinespace
8B natural GSM8K/ARC-Challenge cells: one seed's grids hit a
deterministic CUDA out-of-memory failure on their L40S venue
(long-generation zero arm, twice at the same progress point) &
None: no seed carries an incomplete arm grid &
Seed replaced wholesale by the next integer seed (46 $\to$ 47) per
the replacement rule recorded in advance, before any unblinding; the
replaced seed's partial outputs are preserved unread \\
\addlinespace
MedQA cell hit the same failure class in its smoke &
None: failure preceded production &
Cell re-venued to A100-SXM4-80GB before production \\
\addlinespace
phi-4 calibrated cell: first smoke failed its arm-identity check ---
the run-identity stamp recorded the cache intervention as null in
every arm &
None: per-example intervention events and logs show the
interventions executed as configured; the defect was in the metadata
stamping path, not in execution &
Stamping code fixed to record the executed intervention; rerun smoke
passed; production ran on the fixed revision; the failed smoke is
preserved in the release \\
\addlinespace
C2C integration required five staged attempts before a clean
production run &
None: only the clean production run is reported &
Staged attempts disclosed (platform failure, memory-gate mismatch,
artifact layout, fusion dtype, success) \\
\addlinespace
KVComm port required a calibration chain and a mid-cell dispatch
change (serial to four-lane fan-out) &
None: same jobs, same result root; selection fixed before production
(11 effective layers) &
Per-job GPU recorded; calibration chain disclosed \\
\bottomrule
\end{tabular}
\caption{Run deviations and staged-integration events with their
effect on the reported estimands and their resolutions. Rules cited
as ``recorded in advance'' were written before the corresponding
outcomes existed.}
\label{tab:deviations}
\end{table}

\subsection{Released Artifacts}

The release comprises: the audit harness and the
calibration-instrument generator (registry and prose-carrier
construction, control-arm builders, and the per-example token
audit); per-cell run manifests with per-arm raw outcome ledgers,
recording model checkpoint revision, software stack, physical GPU,
batch geometry, and seeds; each cell's dated design document --- the
arm set, validity conditions, and analysis rule as recorded before
that cell's outcomes existed; the failed smokes and superseded runs
of Table~\ref{tab:deviations}; and the Llama-3.2-3B diagnostic probe
scripts with their per-run ledgers and audit logs. All reported
numbers are recomputed from these raw per-example records rather
than from in-run summaries.

\section{Extended Related Work}
\label{app:related}

\paragraph{Latent communication designs.}
Beyond the three audited systems, related designs relay activations or
hidden ``thoughts'' directly
\citep{ramesh2025_communicating,du2025_enabling,zheng2025_thought},
hybridize latent and text protocols \citep{mou2026_hylat}, extend the
paradigm to memory and multimodal settings
\citep{fu2026_latentmem,liu2026_vision,chen2026_see}, package
reusable KV-communication modules for broader multi-agent
architectures \citep{jin2026_agent}, and reduce relay cost while
preserving reported gains \citep{li2026_when}, including quantized
cache handoff for on-device deployment \citep{honavar2026_qkvshare}
--- a regime where compression may act on the generic and the
example-specific cache contributions separately, exactly the two
terms our decomposition reports. Application-domain designs bring
genuine information asymmetry: in cross-institution medical
diagnosis, hospital agents keep private records local and relay
compact latent KV blocks, with information leakage an explicit
design concern \citep{wang2026_medlatentdx} --- a deployed analogue
of the sender-private construction audited here and a natural
setting for realistic private-information audits beyond synthetic
registries. At the far end, communication itself becomes a training
target: end-to-end optimization of KV-mediated protocols
\citep{yu2026_learning} produces learned channels rather than
delivered fixed mechanisms, and such systems are the natural next
audit target --- a protocol trained for communication is where
example-specific pairing value would most plausibly appear, and
our instrument applies unchanged. See
\citet{liu2026_beyond,chen2026_five} for surveys. The lineage runs
back to learned inter-agent channels in multi-agent RL and emergent
communication
\citep{foerster2016_learning,sukhbaatar2016_learning,lazaridou2016_multi,mordatch2017_emergence}.

\paragraph{Mismatched-latent controls in prior work.}
The nearest methodological precedent for our content control is
Interlat \citep{du2025_enabling}, which pairs matched and mismatched
latent messages both as a training objective --- a separation loss
that teaches the receiving agent to distinguish task-matched from
cross-task latents --- and as a diagnostic battery, including
cross-task mismatch, covariance-matched Gaussian surrogates, and
random rotations. The audit here differs on three axes. First,
Interlat's mismatch sensitivity is trained in and then diagnosed on
its own system; we measure mismatch sensitivity post hoc on released
systems never optimized for it, which is what makes the result an
audit of delivered artifacts rather than a property of a training
recipe. Second, Interlat's perturbations act on last-hidden-state
messages with cross-task mismatch, which carries distribution shift;
our derangement operates at the KV-cache relay boundary, within
batch, and is marginal-preserving, so the contrast isolates pairing
from shift (Appendix~\ref{app:donor-similarity}). Third,
the audit adds the positive calibration that separates channel
failure from receiver redundancy, and equivalence-tested natural
cells. Mismatch-based probing of internal-state transfer has
otherwise produced both positive and negative cross-model findings
\citep{oozeer2025_activation,chen2025_transferring,zhang2026_negative};
the negative findings are consistent with our C2C reading --- a
delivered cross-model projector with no detected example-specific
transfer under an information-asymmetry probe --- while our
positive calibrated cells show the same instrument detects transfer
where it exists.

\paragraph{Latent reasoning.}
Latent (continuous, non-verbalized) reasoning replaces intermediate
token generation with hidden-state recycling within a single model
\citep{hao2024_training}. LatentMAS combines latent reasoning with
latent communication, and its own ablation attributes gains to each
component separately; our audit isolates the communication component
by holding the reasoning machinery fixed within every arm contrast.

\paragraph{Cache- and activation-level interventions.}
KV-cache editing serves efficiency (quantization, eviction
\citep{an2026_rest,bui2026_make}) and safety auditing of shared caches
\citep{asif2026_lcguard,brito2026_when,wang2026_out}. The derangement
arm is an inter-\emph{agent} analogue of activation patching /
causal tracing
\citep{meng2022_locating,wang2022_interpretability,zhang2023_towards,geiger2021_causal},
extending the patching logic from model internals to the
communication boundary between agents while preserving the cache
distribution.

\end{document}